\documentclass[journal,comsoc]{IEEEtran}

\usepackage{amssymb,color}
\usepackage{amsmath}
\usepackage{bbm}
\usepackage{graphicx}
\usepackage{epstopdf}
\usepackage{amsthm,amsfonts}
\usepackage{subcaption}
\usepackage{setspace}
\usepackage{lipsum}
\usepackage{multicol}

\usepackage{accents}

\usepackage{algorithm}
\usepackage{algorithmic}
\usepackage{psfrag}
\usepackage{cite}
\usepackage{pifont}
\usepackage[font={small}]{caption}

\allowdisplaybreaks

\makeatletter
\newcommand{\oset}[3][0ex]{%
  \mathrel{\mathop{#3}\limits^{
    \vbox to#1{\kern-2\ex@
    \hbox{$\scriptstyle#2$}\vss}}}}
\makeatother

\begin{document}
\title{Partial Non-Orthogonal Multiple Access (NOMA) in Downlink Poisson Networks} 

\author{
\IEEEauthorblockN{\large Konpal Shaukat Ali$^{\dagger}$, Ekram Hossain$^{\dagger}$, and Md. Jahangir Hossain$^{+}$}

\thanks{
$^{\dagger}$ The authors are with the Department of Electrical and Computer Engineering, University of Manitoba, Winnipeg, Canada (Email: \{konpal.ali, ekram.hossain\}@umanitoba.ca). E. Hossain is the corresponding author.

$^+$ The author is with the School of Engineering, the University of British Columbia (Okanagan Campus), Canada (Email:  jahangir.hossain@ubc.ca). 
}
}

\maketitle

\begin{abstract} 
Non-orthogonal multiple access (NOMA) allows users sharing a resource-block to efficiently reuse spectrum and improve cell sum rate $\mathcal{R}_{\rm tot}$ at the expense of increased interference. Orthogonal multiple access (OMA), on the other hand, guarantees higher coverage. We introduce partial-NOMA in a large two-user downlink network to provide both throughput and reliability. The associated partial overlap controls interference while still offering spectrum reuse. The nature of the partial overlap also allows us to employ receive-filtering to further suppress interference. For signal decoding in our partial-NOMA setup, we propose a new technique called flexible successive interference cancellation (FSIC) decoding. We plot the rate region abstraction and compare with OMA and NOMA. We formulate a problem to maximize $\mathcal{R}_{\rm tot}$ constrained to a minimum throughput requirement for each user and propose an algorithm to find a feasible resource allocation efficiently. Our results show that partial-NOMA allows greater flexibility in terms of performance. Partial-NOMA can also serve users that NOMA cannot. We also show that with appropriate parameter selection and resource allocation, partial-NOMA can outperform NOMA.

\end{abstract}


\section{Introduction}

{Traditionally, users (UEs) avoid interference from other UEs being served by the same base station (BS) by a multiple access technique known as orthogonal multiple access (OMA) which allocates orthogonal resources to these UEs. This is done by allotting UEs different time slots or different frequency channels; in fact, the the available time and frequency resources are split into a grid of what are referred to as time-frequency resource-blocks (also referred to as resource-blocks from hereon). This way a UE in one resource-block has orthogonal resources to a UE in any other resource-block. In contrast to OMA, in non-orthogonal multiple access (NOMA), multiple UEs share a time-frequency resource-block for transmissions by superposing their messages in the power domain, i.e., multiple UEs transmit their messages in the same time slot over the same frequency channel using different power levels for their messages. Hence, NOMA UEs improve spectral reuse by sharing a resource-block} with other UEs but the price paid is the introduction of interference with UEs being served by the BS on the same resource-block, i.e., intracell interference. 

NOMA employs successive interference cancellation (SIC) for the decoding of these superposed messages. {SIC requires ordering of NOMA UEs based on some measure of channel strength. Most of the existing works on NOMA in the literature order UEs based on the mean signal power received \cite{N2,N13,N4,N6,myNOMA_icc,myNOMA_meta} or on the quality of the transmission channel such as the fading coefficient \cite{N5, N7, N8,N9}, the fading-to-noise ratio \cite{N15}, the instantaneous received signal-to-intercell-interference-and-noise ratio \cite{myNOMA_tcom}, and the instantaneous received signal-to-intercell-interference ratio \cite{N16}.\footnote{{Note that UE ordering in NOMA is based on a measure of channel strength and does not imply that the message of a strong (weak) UE has high (low) transmission power.}} Such UE ordering and appropriate resource allocation allows a UE to decode messages of UEs weaker than itself and treat the messages of UEs stronger than itself as noise. It has been shown extensively in the literature that NOMA is superior to OMA in terms of the sum throughput that can be achieved \cite{N1,N8,N9,N13,N14,myNOMA_tcom,N15,N6,N18,N16,myNOMA_icc}.

While NOMA allows complete sharing of a resource-block, {thereby resulting in better throughput, the introduction of intracell interference deteriorates the coverage of NOMA UEs. OMA, on the other hand, has no intracell interference, and therefore, provides superior coverage}. However, OMA limits only one UE to a resource-block thereby not making efficient use of the scarce spectrum {which leads to lower throughput. To cater to better coverage requirements than NOMA as well as better throughput requirements than OMA}, we introduce the concept of partial-NOMA. In partial-NOMA, UEs share only a fraction of the resource-block, this way we can {reduce the intracell interference but still allow some spectrum reuse}. The motivation behind this concept is to introduce flexibility into the system by having control over how much of the resource-block overlaps and does not overlap for each UE. This way, partial-NOMA is a general technique which offers flexibility between the two extremes of traditional OMA and NOMA.

To the best of our knowledge, only the work in \cite{pnoma} studies a partial-NOMA like setup in a two-user downlink scenario. Different from our work, the authors study a single-cell setup where the non-overlapping areas of the resource-block allow each UE's message to be given full power while the overlapping regions share power between the two UEs so that it sums to the full power (\cite[Fig. 1]{pnoma}). Such an analysis is equivalent to studying the average performance of a system with OMA in some resource-blocks and NOMA in other resource-blocks. In our work, we study the performance of shared power in one resource-block that sums up to the total power, with a partial area of overlap (c.f. Fig. \ref{alphaBetaNOMA}). Additionally, we study a large network since a single-cell setup does not account for intercell interference. Accounting for intercell interference is particularly important for setups where multiple UEs share a resource-block (such as NOMA and partial-NOMA) as it has a drastic negative impact on both performance and resource allocation, as was shown in \cite{my_nomaMag}. Stochastic geometry has succeeded in providing a unified mathematical paradigm for modeling large wireless cellular networks and characterizing their operation while taking into account intercell interference \cite{MH_Book2,3B1,h_tut,di_renzo}. Works on NOMA such as \cite{N6,N18,N16,myNOMA_meta,myNOMA_icc,distRanking} use stochastic geometry based modeling for analyzing large networks. {A number of works on NOMA have focused on resource allocation as well \cite{N15,N9,N13,N14,FD_N4,myNOMA_tcom,myNOMA_icc,mankarNOMA}. We also study resource allocation in this paper for a partial-NOMA setup.}

In this work, we analytically study a large multi-cell downlink system that employs partial-NOMA for the two-user setup. {Partial sharing of a resource-block in our work is accomplished by having the two signals overlap only with a fraction of each other in the frequency domain while having complete access to the entire time slot. An overlap in the frequency domain allows us to} employ matched filtering at the receiver side to reduce the impact of the intracell interference\footnote{Note that the intracell interference is only a fraction of the intracell interference experienced in the case of traditional NOMA. However, with matched filtering at the receiver we can further reduce its impact.}. {Traditional NOMA relies on SIC, which involves a strong UE decoding and removing the message of the weak UE before it decodes its own message.} In partial-NOMA, however, after matched filtering at the receiver side, the message of the weak UE may be too weak for the strong UE to be able to decode. To combat this issue and improve performance, we propose a new decoding scheme, which we call flexible-SIC (FSIC) decoding. Using stochastic geometry tools, we mathematically analyze the performance of a large network with partial-NOMA employing FSIC decoding. The rate region for NOMA and OMA has been studied in the literature. To benchmark the performance of a partial-NOMA setup, we study the rate region abstraction for different values of the overlap\footnote{The rate region will be abstracted into two figures to make it easier to read for different values of the overlap.}. The partial-NOMA setup introduces two new parameters that can be varied, the overlap and the amount of non-overlap given to each UE. Accordingly, the rate region for different values of the overlap is not enough to quantify performance. A more practical problem is that of maximizing cell sum rate, defined as the sum of the throughput of the two UEs sharing a resource-block, subject to a {\em threshold minimum throughput} (TMT) constraint on the individual UEs. In this context, we formulate an optimization problem and propose an algorithm for resource allocation. We show a significant reduction in complexity between our proposed algorithm and an exhaustive search.

The contributions of this paper can be summarized as follows:
\begin{itemize}

\item We show that our partial-NOMA setup can result in lower intercell interference than both traditional OMA and NOMA thanks to the received filtering.

\item While it is well known that the impact of bandwidth is more significant on throughput than the impact of interference, we show that partial-NOMA is able to outperform traditional NOMA in terms of the cell sum rate in the rate region. This superiority is due to the associated received filtering and proposed FSIC decoding.

\item We show that in terms of individual UE throughput, NOMA that has complete overlap is closer to OMA that has no overlap, while partial-NOMA allows the individual rates to stray farther away. This highlights the greater flexibility in individual UE performance that a partial-NOMA setup introduces.
\item We show that while traditional NOMA cannot support UEs with high transmission rate requirements, partial-NOMA can. Thus instead of allocating an entire resource-block to such UEs, they can be served in a partial-NOMA fashion to efficiently reuse the spectrum.
\item We show that with {careful resource allocation, partial-NOMA with a range of overlap values outperforms traditional NOMA, in terms of cell sum rate. We also show that an optimum overlap exists that maximizes the cell sum rate given a threshold minimum throughput constraint.}
\end{itemize}



{The rest of the paper is organized as follows. The system model is described Section II. The SINR analysis, FSIC decoding and relevant statistics are in Section III. In Section IV, the rate region for partial-NOMA is studied, an optimization problem is formulated, and an algorithm is proposed to solve the problem. The results are presented in Section V and the paper is concluded in Section VI.}

\textit{Notation:} Vectors are denoted using bold text, $\|\textbf{x} \|$ denotes the Euclidean norm of the vector $\textbf{x}$, {$b(\textbf{x},r)$ denotes a disk centred at $\textbf{x}$ with radius $r$}. $\mathcal{L}_X(s)=\mathbb{E}[e^{-sX}]$ denotes the Laplace transform (LT) of the PDF of the random variable $X$. We use the indicator function, denoted as $\mathbbm{1}_A$, to have value 1 when event $A$ occurs and to be 0 otherwise. The ordinary hypergeometric function is denoted by ${}_2F_1$. {We use $\rm{Sinc}(x)=\sin(\pi x)/(\pi x)$ when $x \neq 0$, and $\rm{Sinc}(x)=1$ when $x=0$.}
%

\section{System Model}\label{SysMod}

\subsection{Network Model for Partial-NOMA}

We consider a downlink cellular network where each BS serves two UEs in each time-frequency resource-block. In traditional NOMA, the two UEs share the entire resource-block {(i.e., each UE has access to the full time slot and frequency channel associated with the resource-block)}, having their messages multiplexed in the power domain. In contrast to this, in partial-NOMA, the messages of the two UEs only overlap over a fraction $\alpha$ of the resource-block. {This overlap of the resource-block can be achieved in two ways:
\begin{enumerate}
\item Both UEs transmit over the entire time slot but there is only an overlap of the fraction $\alpha$ of the frequency channel shared by the two UEs.
\item Both UEs occupy the entire frequency channel but simultaneous transmission only happens for a fraction $\alpha$ of the time slot.  
\end{enumerate}
As mentioned in Section I, in this work, we consider the first approach, i.e., the two UEs only share a fraction $\alpha$ of the frequency resources while transmitting in the entire time slot of a resource-block. This allows us to take advantage of matched filtering explained in Section II-B which would not have been possible with the second approach. With a slight abuse of notation, in the remainder of the manuscript, an overlap of $\alpha$ of the resource-block refers to an overlap of $\alpha$ in the frequency domain. Since the entire time slot is available to both UEs, we disregard this aspect when referencing the partial overlap of the resource-block. Note that we stick to the notation of resource-block to remain consistent with works on traditional NOMA.}

 As shown in Fig. \ref{alphaBetaNOMA}, we refer to the fraction of the resource-block designated to only $\text{UE}_1$ by $\beta$; consequently, the fraction designated to only $\text{UE}_2$ is $1-\alpha-\beta$. Thus, the effective bandwidth of $\text{UE}_1$ is $\text{BW}_1=\alpha+\beta$, and that of $\text{UE}_2$ is $\text{BW}_2=1-\beta$. The BSs use fixed-rate transmissions where the transmission rate of each UE can be different. Such transmissions result in effective rates, {referred to as the throughput of the UEs}, that are lower than the transmission rate because of outage.

\begin{figure}[htb]
\begin{minipage}[htb]{0.97\linewidth}
\centering\includegraphics[scale=0.4]{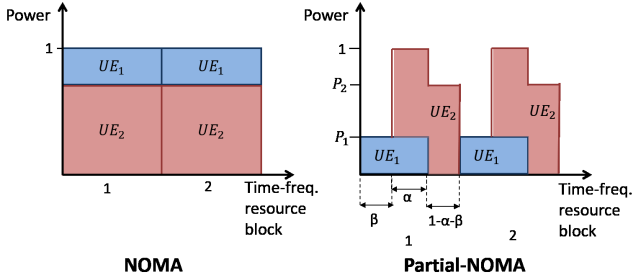}
\caption{A contrast between resource allocation in traditional NOMA and partial-NOMA.}\label{alphaBetaNOMA} 
\end{minipage}\;\;\;\;\;\;
\end{figure}

The BSs are distributed according to a homogeneous Poisson point process (PPP) $\Phi$ with intensity $\lambda$. To the network we add a BS at the origin $\textbf{o}$, which under expectation over $\Phi$, becomes the typical BS serving UEs in the typical cell. We study the typical cell in the remainder of this work. Note that as $\Phi$ does not include the BS at $\textbf{o}$, the set of interfering BSs for the UEs in the typical cell is denoted by $\Phi$. We denote by $\rho$ the distance between the typical BS at $\textbf{o}$ and its nearest neighboring BS. Since $\Phi$ is a PPP, the distribution of the distance $\rho$ thus follows
\begin{align}
f_{\rho}(x)=2\pi \lambda x e^{-\pi \lambda x^2}, \;\;\; x\geq 0.\label{f_rho}
\end{align} 

An important challenge associated with NOMA is the selection of the UEs that share a resource-block, which is referred to as UE clustering. There is a common misconception that clustering NOMA UEs with high channel disparity is beneficial. However, works such as \cite{cry1} have shown that this not necessarily the case and in \cite{myNOMA_meta,myNOMA_channDisp} it is explicitly shown that in fact clustering UEs with lower channel disparity but overall better channel conditions is superior. Hence, we use the cell-center model for the selection of NOMA UEs employed in \cite[Model 1]{myNOMA_tcom}. As such, we consider a disk centred at {\textbf{o}} with radius $\rho/2$, $b(\textbf{o},\rho/2) $, referred to as the in-disk. The in-disk is the largest disk centred at the serving BS that fits inside the Voronoi cell. {The rationale behind employing this model is that UEs outside of this disk are relatively far from their BS, have weaker channels and thus are better served in their own resource-block (without sharing) or even using {coordinated multipoint (CoMP) transmission} if they are near the cell edge \cite{comp1,comp2}. These UEs are not discussed further in this work. We focus on UEs inside the in-disk since they have good channel conditions, yet enough disparity among themselves, and thus can effectively be served} while (partially) sharing a resource-block. The two UEs are located uniformly at random in the in-disk, independent of one another. {Note that this way the UEs form a Poisson cluster process where two daughter points are placed independently and uniformly at random on disks of varying random radii. Such a model is superior to using a Matern cluster process where the radius of the disks is fixed and thus: 1) risks a disk of one BS going into the cell of a neighboring BS, 2) risks overlap between disks.} 


A BS transmits with a power budget $P$, where the power for the signal intended for $\text{UE}_i$ is denoted by $P_i$ and $P=\sum_{i=1}^2 P_i$; without loss of generality we set $P=1$. A Rayleigh fading environment is assumed such that the fading coefficients are i.i.d. with a unit mean exponential distribution. A power law path-loss model is considered where the signal power decays with distance $r$ at the rate $r^{-\eta}$, where $\eta>2$ denotes the path-loss exponent.

{As has been mentioned, NOMA requires ordering the UEs based on some measure of channel strength. In this work, we order the UEs based on the link distance, $R$, between the typical BS at $\textbf{o}$ and its UE uniformly distributed in the in-disk is conditioned on $\rho$. Note that ordering based on increasing link distance is equivalent to ordering based on the decreasing mean signal power received, i.e., $R^{-\eta}$.} We thus refer to the strong UE, with the shorter link distance, as $\text{UE}_1$ and the weak UE as $\text{UE}_2$. As the order of the UEs is known at the BS, we use ordered statistics for the pdf of $R_i$, the ordered link distance of $\text{UE}_i$, where $i \in \{1,2\}$. Hence, {using the theory of order statistics \cite{N16_18}}, in the typical cell
\begin{align}
f_{R_i \mid \rho}(r \mid \rho) = \frac{16 r}{\rho^2} \left( \frac{4r^2}{\rho^2} \right)^{i-1} \left( 1- \frac{4r^2}{\rho^2} \right)^{2-i} \;\;\; 0\leq r \leq \frac{\rho}{2}.
\end{align}
Note that as there is no interfering BS inside $b(\textbf{o},\rho) $, the nearest interfering BS from $\text{UE}_i$ is at least $\rho-R_i$ away. Thus the in-disk model allows a larger lower bound on the distance from the nearest interferer than the usual lower bound of link distance for UEs in a downlink Poisson network \cite{myNOMA_tcom}.

\subsection{Filtering at the Receiver}
Since partial-NOMA is studied in this work, the signal of interest at a UE only overlaps with a fraction $\alpha$ in the resource-block with the other UE's signal. Accordingly, it is important to consider the impact of matched filtering at the receiver, along the lines of the works in \cite{AlAmmouri,tsikyJrnl}. The message of the signal for $\text{UE}_i$ is transmitted using the pulse shape $s_i(t) \leftrightarrow S_i(f)$. The signal is passed through a matched filter $H_i(f)=S_i^*(f)$ at the receiver before decoding, where $S_i^*(f)$ denotes the conjugate of $S_i(f)$. Accordingly, the impact of the interference caused at $\text{UE}_i$ by a message that has an $\alpha$-overlap, is referred to as the effective interference factor $\mathcal{I}_i(\alpha, \beta)$. Note that $\mathcal{I}_i(\alpha, \beta)$ is a function of both $\alpha$ and $\beta$ as both are required to determine the center frequencies of the signals. In this work, we assume that square pulses are used by both UEs for transmissions. Because of the choice of $H_i(f)$ and since the same pulse shape is transmitted by both UEs, $\mathcal{I}_i(\alpha, \beta)=\mathcal{I}_j(\alpha, \beta)$ when $i \neq j$; hence, in the remainder of the manuscript, we drop the subscript. {Along the lines of \cite{AlAmmouri},} the effective interference factor as a function of $\beta$ and the overlap $\alpha$ is calculated as follows:
\begin{align}
\mathcal{I}(\alpha, \beta)= \left( \int\limits_{\beta}^{\beta+\alpha} \frac{\rm{Sinc}\left( \frac{2(f-f_a)}{\text{BW}_1}\right)}{E_1}  \frac{\rm{Sinc}\left( \frac{2(f-f_b)}{\text{BW}_2}\right)}{E_2} df \right)^{{2}},
\end{align}
where the center frequency of $\text{UE}_1$'s message is $f_a=\frac{\alpha+\beta}{2}$ and $\text{UE}_2$'s message is $f_b=\frac{1+\beta}{2}$. The factors $E_i$ for $i \in \{1,2\}$ are used to scale the energy to 1 and are calculated as $E_i^2= \int_{-\text{BW}_i/2}^{\text{BW}_i/2} \rm{Sinc}^2 \left( \frac{2f}{\text{BW}_i}\right) df$. 

Note that we define $\mathcal{I}(\alpha, \beta)$ scaled to unit power from the interfering message as this is the fraction of the interference that actually impacts the received message. The actual interfering power, which is the transmit power of the message scaled by $\mathcal{I}(\alpha, \beta)$, will be accounted for in the SINR analysis. Additionally, it is important to emphasize that any interfering message that has an $\alpha$-overlap with the message of $\text{UE}_i$, whether it is from inside the typical cell or outside, will have its power scaled by $\mathcal{I}(\alpha, \beta)$.


\begin{figure}[htb]
\begin{minipage}[htb]{0.97\linewidth}
\centering\includegraphics[width=0.9\columnwidth]{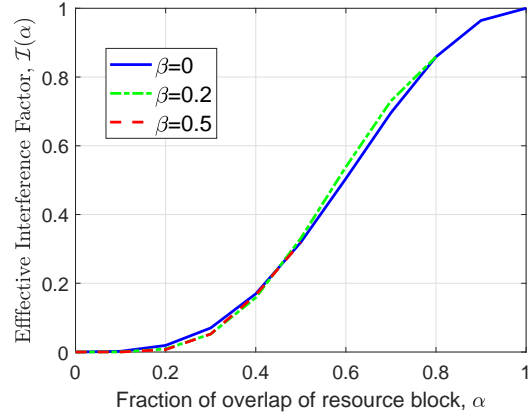}
\subcaption{As a function of the overlap $\alpha$ for different values of $\beta$.}\label{intrfFactorVsAlpha}
\end{minipage}
\begin{minipage}[htb]{0.97\linewidth}
\centering\includegraphics[width=0.9\columnwidth]{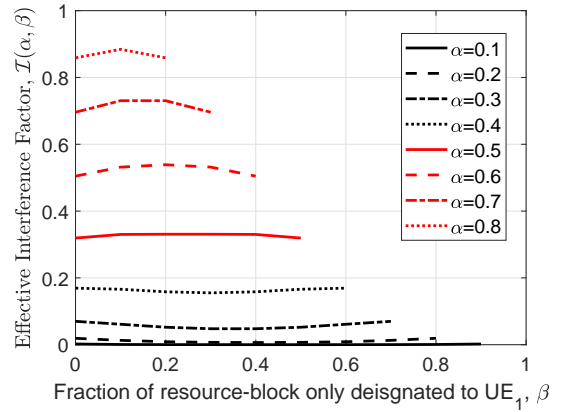}
\subcaption{As a function of $\beta$ for different values of the overlap $\alpha$.}\label{intrfFactorVsBeta}
\end{minipage}
\caption{Effective interference factor for square pulses.}\label{intrfFactor}
\end{figure}

{Fig. \ref{intrfFactor} plots $\mathcal{I}(\alpha, \beta)$ for the case of square pulses used in this work. It should be noted from Fig. \ref{intrfFactorVsAlpha} that when the overlap $\alpha$ is 0 (1), as in the case of OMA (traditional NOMA), the effective interference factor is 0 (1). Also note that for a given $\beta$, $\alpha$ is restricted to a maximum value of $1-\beta$; hence, when $\beta>0$, $\alpha \leq 1-\beta <1$. Accordingly, $\mathcal{I}(\alpha, \beta)$ in Fig. \ref{intrfFactorVsAlpha} is plotted up to the maximum value of the overlap $\alpha$ possible. For a given $\alpha$, on the other hand, $\beta$ can take on values $0 \leq \beta \leq \beta_{\rm max}$, where $\beta_{\rm max}=1-\alpha$. We observe in Fig. \ref{intrfFactorVsBeta} that for each value of $\alpha$, the curve for $\mathcal{I}(\alpha,\beta)$ vs. $\beta$ is symmetric about the midpoint which occurs at $\beta=\beta_{\rm max}/2$. Additionally, it is important to note that for smaller (larger) values of $\alpha$, $\mathcal{I}(\alpha,\beta)$ decreases (increases) from $\beta=0$ to $(1-\alpha)/2$ (i.e., $\beta_{\rm max}/2$ the point of symmetry) and then increases (decreases) from $\beta=(1-\alpha)/2$ to $1-\alpha$.}

\subsection{Decoding of Partial-NOMA}
Traditional NOMA employs successive interference cancellation (SIC) for decoding. Hence, in the downlink, the strong UE first decodes the weak UE's message and subtracts it from the signal before decoding its own message. The weak UE, on the other hand, only decodes its own message, treating the signal of the strong UE as noise. Accordingly, the power allocation and transmission rate are designed so that decoding the weak UE's message is easier; this is done by allocating it higher power than the strong UE and/or low transmission rate. Unlike traditional NOMA, however, the intracell interference is scaled by the effective interference factor $\mathcal{I}(\alpha, \beta)$ which can be much lower than 1. While this is always beneficial for the weak UE which treats intracell interference as noise, it can make traditional SIC decoding difficult for the strong UE which has to decode the weak UE's message as the effective power of the weak message at the strong UE could be significantly lower. In this context, we propose FSIC in the next section which is a more effective decoding technique for the partial-NOMA setup. 

\section{SINR Analysis}

\subsection{The SINRs of Interest}
In the two-user downlink partial-NOMA setup, we are interested in the following SINRs associated with decoding:
\begin{itemize}
\item the message of $\text{UE}_2$ at $\text{UE}_2$: ${\rm SINR}_2^2$,
\item the message of $\text{UE}_2$ at $\text{UE}_1$: ${\rm SINR}_2^1$,
\item the message of $\text{UE}_1$ at $\text{UE}_1$ after the message of $\text{UE}_2$ has been removed: ${\rm SINR}_1^1$,
\item the message of $\text{UE}_1$ at $\text{UE}_1$ if the message of $\text{UE}_2$ is not removed: ${\widetilde{\rm SINR}}_1^1$.
\end{itemize}
Accordingly, the SINR of decoding the $j^{th}$ message at $\text{UE}_i$, where $j \geq i$, assuming that messages of all UEs weaker than $\text{UE}_j$ have been removed is
\begin{align}
{\rm SINR}_j^i \! =\! \frac{h_i R_i^{-\eta} \left(P_j \mathcal{I}(\alpha, \beta) \mathbbm{1}_{i\neq j} +  P_i\mathbbm{1}_{i= j} \right)}{h_i R_i^{-\eta} \!\left( \left(\!1 \!-\!\sum_{k=j}^2 P_k \! \right) \mathcal{I}(\alpha, \beta) \mathbbm{1}_{i=j} +  P_i \mathbbm{1}_{i \neq j} \right) \!+\! \tilde{I}^{\rm\o}_i \! + \!\sigma^2}. \label{sinr_ij}
\end{align}

The intercell interference experienced at $\text{UE}_i$ is $\tilde{I}^{\rm\o}_i= \left(P_i +(1-P_i) \mathcal{I}(\alpha, \beta) \right)\sum_{\textbf{x} \in \Phi}  g_{\textbf{y}_i} {\|\textbf{y}_i\|}^{-\eta} $, where $\textbf{y}_i=\textbf{x}-\textbf{u}_i$, $\textbf{u}_i$ is the location of $\text{UE}_i$. The fading coefficient from the serving BS (interfering BS) located at $\textbf{o}$ ($\textbf{x}$) to $\text{UE}_i$ is $h_i$ ($g_{\textbf{y}_i}$). For notational convenience, we define the intercell interference scaled to unit transmission power by each interferer as $I^{\rm\o}_i$; hence, $\tilde{I}^{\rm\o}_i= \left(P_i + (1-P_i) \mathcal{I}(\alpha, \beta) \right) I^{\rm\o}_i$ . The noise power is denoted by $\sigma^2$. Using these notations, $\rm{\widetilde{SINR}}_1^1$ is
\begin{align}
{\rm \widetilde{SINR}}_1^1 \! =\! \frac{h_1 R_1^{-\eta} P_1 }{h_1 R_1^{-\eta} P_2 \mathcal{I}(\alpha, \beta)+ \tilde{I}^{\rm\o}_1 +\sigma^2}. \label{sinr1_new}
\end{align}
\textbf{\emph{Remark 1:}} It ought to be mentioned that because of the received filtering and associated interference factor $\mathcal{I}(\alpha,\beta)$, the partial-NOMA setup experiences lower intercell interference than both OMA and traditional NOMA.

\subsection{FSIC Decoding}

FSIC decoding for $\text{UE}_2$ is the same as that of SIC decoding, i.e., $\text{UE}_2$ decodes its message while treating the interference from the message of $\text{UE}_1$ as noise. Accordingly, the event of successful decoding at $\text{UE}_2$ is defined as $C_2= \left\lbrace {\rm SINR}_2^2>\theta_2 \right\rbrace$, where $\theta_2$ is the SINR threshold corresponding to the transmission rate of $\text{UE}_2$'s message $\log(1+\theta_2)$. Note that the (intracell) interference in the case of partial-NOMA is lower than in the case of traditional NOMA as it is scaled by $\mathcal{I}(\alpha, \beta)$. For $\text{UE}_1$, on the other hand, the message of interest can be decoded successfully if either: 1) the message of $\text{UE}_2$ can be decoded (treating the message of $\text{UE}_1$ as noise) and removed, followed by decoding of the message of $\text{UE}_1${\footnote{{We assume perfect SIC in this work. As a result, there is no leakage of the canceled message of the weak UE which would add as interference when decoding the message of the strong UE.}}}, or 2) the message of $\text{UE}_1$ can be decoded while treating the interference from the message of $\text{UE}_2$ as noise. Accordingly, it is defined by the following joint event $C_1= \left\lbrace \left({\rm SINR}_2^1 > \theta_2 \cap {\rm SINR}_1^1 > \theta_1 \right) \cup {\rm \widetilde{SINR}}_1^1>\theta_1 \right\rbrace$, where $\theta_1$ is the SINR threshold corresponding to the transmission rate of $\text{UE}_1$'s message $\log(1+\theta_1)$.

\subsection{Coverage Analysis and Throughput}
\textbf{\emph{Lemma 1:}} The event $C_2$ can be rewritten as follows:
\begin{align}
C_2= h_2  >  R_2^{\eta} (\tilde{I}^{\rm\o}_2 + \sigma^2) \bar{M}_2 \label{C2} 
\end{align}
where we have used 
\begin{align*}
\tilde{P}_2=P_2 - \theta_2 P_1 \mathcal{I}(\alpha, \beta),
\end{align*}
such that
\begin{align}
\bar{M}_2= \frac{\theta_2}{\tilde{P}_2}. \label{Mbar_2}
\end{align}
\begin{IEEEproof}
Using \eqref{sinr_ij} and the definition of $C_2$, \eqref{C2} is obtained as follows:
{\begin{align*}
C_2 &= \left\lbrace \mathrm{SINR}_2^2>\theta_2 \right\rbrace \\
 &= \left\lbrace \frac{h_2 R_2^{-\eta} P_2 }{h_2 R_2^{-\eta} P_1 \mathcal{I}(\alpha,\beta) + \tilde{I}^{\mathrm\o}_2 +\sigma^2 }  >\theta_2 \right\rbrace \\
 &= \left\lbrace h_2    > R_2^{\eta} \left(\tilde{I}^{\mathrm\o}_2 + \sigma^2 \right) \frac{\theta_2 }{{{P_2  - \theta_2 P_1 \mathcal{I}(\alpha,\beta)}} }  \right\rbrace.
 \end{align*}}
\end{IEEEproof}

\textbf{\emph{Lemma 2:}} The event $C_1$ can be rewritten as follows:
\begin{align}
C_1= h_1  >  R_1^{\eta} (\tilde{I}^{\rm\o}_1 + \sigma^2) \bar{M}_1 . \label{C1}
\end{align}
where we have used \begin{align*}
M_0=\frac{\theta_1}{ \tilde{P}_1},\\
M_1= \max \left\lbrace \frac{\theta_2}{ \tilde{P}_2^1 }, \frac{\theta_1}{P_1} \right\rbrace,\\
\tilde{P}_1= P_1 - \theta_1 P_2 \mathcal{I}(\alpha, \beta),\\
\text{and } \tilde{P}_2^1= P_2 \mathcal{I}(\alpha, \beta) - \theta_2 P_1,
\end{align*} 
such that
\begin{align}
  \bar{M}_1  =  & \min \left\lbrace M_0  , M_1  \right\rbrace \mathbbm{1}_{\tilde{P}_1>0}  \mathbbm{1}_{\tilde{P}_2^1>0} \mathbbm{1}_{P_1>0}    \; + \;  \nonumber \\
  & M_0 \mathbbm{1}_{\tilde{P}_1>0}  \mathbbm{1}_{ \tilde{P}_2^1 \leq 0 \; \cup P_1 \leq 0 } \; + \; M_1 \mathbbm{1}_{\tilde{P}_1 \leq 0} \mathbbm{1}_{\tilde{P}_2^1>0}    \mathbbm{1}_{P_1>0} . \label{Mbar_1}
\end{align} 
\begin{IEEEproof}
Using \eqref{sinr_ij}, \eqref{sinr1_new} and the definition of $C_1$: 
{\begin{align*}
C_1 &=&\left\lbrace \left(\mathrm{SINR}_2^1 > \theta_2 \cap \mathrm{SINR}_1^1 > \theta_1 \right) \cup \mathrm{ \widetilde{SINR}}_1^1>\theta_1 \right\rbrace \\
&=& \Bigg\lbrace \left( \frac{h_1R_1^{-\eta} P_2 \mathcal{I}(\alpha,\beta)}{ h_1R_1^{-\eta} P_1 + \tilde{I}^\mathrm{\o}_1 +\sigma^2  } > \theta_2 \bigcap \frac{h_1R_1^{-\eta} P_1}{ \tilde{I}^\mathrm{\o}_1 +\sigma^2  }  > \theta_1 \right) \bigcup \\
&& \frac{h_1 R_1^{-\eta} P_1 }{h_1 R_1^{-\eta} P_2 \mathcal{I}(\alpha, \beta)+ \tilde{I}^\mathrm{\o}_1 +\sigma^2} >\theta_1 \Bigg\rbrace \\
&=&\left\lbrace  h_1> R_1^{\eta} \left(\tilde{I}^\mathrm{\o}_1 +\sigma^2 \right) M_1   \bigcup   h_1> R_1^{\eta} \left(\tilde{I}^\mathrm{\o}_1 +\sigma^2 \right) M_0 \right\rbrace.
\end{align*}
For the event $\{ \left(\mathrm{SINR}_2^1 > \theta_2 \cap \mathrm{SINR}_1^1 > \theta_1 \right) \}$, $\tilde{P}_2^1 > 0$ and $P_1> 0$ are required. For the event $\{ \mathrm{ \widetilde{SINR}}_1^1>\theta_1 \}$, $\tilde{P}_1 > 0$ is required. Thus, \eqref{C1} is obtained by rewriting $C_1$ in terms of these conditions as follows:
\begin{align*}
C_1=   \Big\{  h_1> & R_1^{\eta} \left(\tilde{I}^\mathrm{\o}_1 +\sigma^2 \right) \Big(  \min\{M_0,M_1 \} \mathbbm{1}_{\tilde{P}_1>0}  \mathbbm{1}_{\tilde{P}_2^1>0} \mathbbm{1}_{P_1>0}     \\
& + M_0 \mathbbm{1}_{\tilde{P}_1>0}  \mathbbm{1}_{ \tilde{P}_2^1 \leq 0 \; \cup P_1 \leq 0 } + M_1 \mathbbm{1}_{\tilde{P}_1 \leq 0} \mathbbm{1}_{\tilde{P}_2^1>0}    \mathbbm{1}_{P_1>0} \Big)  \Big\}  .
\end{align*}
}
\end{IEEEproof}

{
\textbf{\emph{Remark 2:}} Note that the introduction of intracell interference impacts the transmit power of the message being decoded by deteriorating it to what is referred to as the effective transmit power \cite{myNOMA_tcom}. The amount of deterioration is the intracell interference experienced scaled by the SINR threshold corresponding to the transmission rate of the message to be decoded. Thus, $\tilde{P}_2$, $\tilde{P}_1$, and $\tilde{P}_2^1$ are the effective transmit powers that have experienced a reduction from the power of the messages to be decoded ($P_2$, $P_1$ and $P_2 \mathcal{I}(\alpha, \beta)$, respectively).}


Using \cite[Lemma 1]{myNOMA_tcom}, the LT of $I^{\rm\o}_i$ at the typical $\text{UE}_i$ conditioned on $R_i$ and $\rho$, where $u_i=\rho-R_i$, is {approximated as}
\begin{align}
\mathcal{L}_{I^{\rm\o}_i \mid R_i,\rho} (s)  &\!\approx \!  \exp \!\left(\!{   \frac{-2 \pi \lambda s }{(\eta-2){u_i}^{\eta-2}} { }_2F_1  \left( 1,1 \! - \! \delta; 2\! - \! \delta; \frac{-s}{{u_i}^{\eta}} \right)   }\! \right)\!   \frac{1}{1\!+\!s \rho^{-\eta}}  \label{L_I}\\
&\stackrel{\eta=4}= e^{-\pi \lambda \sqrt{s} \tan^{-1} \left(\frac{\sqrt{s}}{u_i^{2}} \right)}  \frac{1}{1+s \rho^{-4}}.
\end{align}

\textbf{\emph{Theorem 1:}} If the effective transmit power is positive, the coverage probability of the typical $\text{UE}_i$ is approximated as
\begin{align}
\mathbb{P} (C_i) \approx  \int\limits_0^{\infty} \! \int\limits_0^{x/2} \! & e^{-r^{\eta} \sigma^2 \bar{M}_i}  \mathcal{L}_{I^{\rm\o}_i \mid R_i,\rho } \left( {r^{\eta}  \bar{M}_i \left(P_i+(1-P_i) \mathcal{I}(\alpha, \beta) \right)} \right)  \nonumber \\
& \times f_{R_i\mid \rho}( r \mid x ) \; dr f_{\rho}(x)\; dx, \label{cvg}
\end{align}
where the LT of $I^{\rm\o}_i$ conditioned on $R_i$ and $\rho$ is approximated in \eqref{L_I}. For $\text{UE}_1$, the effective transmit power is positive if $\{\tilde{P}_2^1>0, P_1>0 \}$ and/or $\tilde{P}_1>0$, while for $\text{UE}_2$, it is positive when $\tilde{P}_2>0$. If effective transmit power is not positive, $\mathbb{P} (C_i)=0$.
\begin{IEEEproof}
{Using $C_i$ in Lemma 1 or Lemma 2, if the effective transmit power is positive, we can write
\begin{align*}
&\mathbb{P}(C_i) = \mathbb{P} \left( h_i > R_i^{\eta} (\tilde{I}^\mathrm{\o}_i + \sigma^2) \bar{M}_i \right) \\
&\stackrel{(a)}=\mathbb{E}_{\rho}  \left[ \mathbb{E}_{R_i \mid \rho} \left[ e^{- R_i^{\eta} \sigma^2 \bar{M}_i}  \mathbb{E}_{I^\mathrm{\o}_i \mid R_i, \rho}  \left[ e^{- R_i^{\eta} \bar{M}_i  \left(P_i + (1-P_i) \mathcal{I}(\alpha, \beta) \right) I^\mathrm{\o}_i } \right] \right]  \right] \\
& \approx \mathbb{E}_{\rho} \! \left[ \mathbb{E}_{R_i \mid \rho} \! \left[ e^{- R_i^{\eta} \sigma^2 \bar{M}_i}  \mathcal{L}_{I^\mathrm{\o}_i \mid R_i, \rho} \! \left( R_i^{\eta} \bar{M}_i  \left(P_i + (1\!-\!P_i) \mathcal{I}(\alpha, \beta) \right)  \right) \right]  \right],
\end{align*}
where (a) follows from $h_i \sim \exp(1)$ and using $\tilde{I}^\mathrm{\o}_i= \left(P_i + (1-P_i) \mathcal{I}(\alpha, \beta) \right) I^\mathrm{\o}_i$. The coverage probability becomes an approximation when the approximate LT of $I^{\rm\o}_i$ (given $R_i$ and $\rho$), $\mathcal{L}_{I^\mathrm{\o}_i \mid R_i, \rho}$, is used. From this, we obtain \eqref{cvg}. Since $h_i \geq 0$, if the effective transmit power is not positive, $\mathbb{P} (C_i)=0$.
}
\end{IEEEproof}

The throughput of $\text{UE}_i$, $i \in \{1,2\}$, for a given SINR threshold of $\theta_i$ corresponding to a transmission rate of $\log(1+\theta_i)$ is given by
\begin{align}
\mathcal{R}_i= \text{BW}_i \; \mathbb{P}(C_i) \log(1+\theta_i). \label{R_i}
\end{align}
We define the cell sum rate $\mathcal{R}_{\rm tot}$ as the sum of the throughput of the UEs in the typical cell; thus, $\mathcal{R}_{\rm tot}=\mathcal{R}_1 + \mathcal{R}_2$.

It ought to be mentioned that in a partial-NOMA setup, the resources to be allocated are the powers, the transmission rates, and the overlapping as well as non-overlapping fractions of the resource-block; this means allocating $P_1 (=1-P_2)$, $\theta_1$, $\theta_2$, $\alpha$ and $\beta$.

\section{Rate Region and Optimization}

\subsection{Rate Region Abstraction}
\begin{figure}
\begin{minipage}[htb]{0.97\linewidth}
\centering\includegraphics[width=0.9\columnwidth]{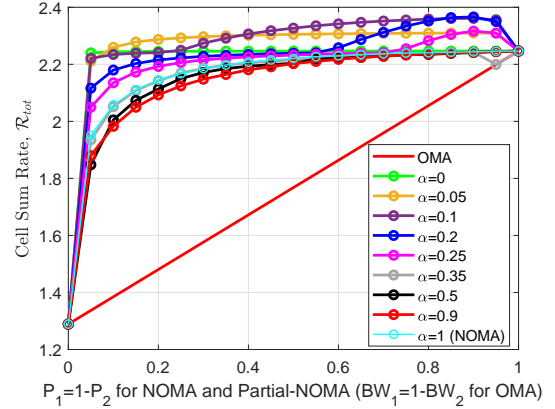}
\subcaption{Cell Sum Rate\\}\label{TotRateVsP}
\end{minipage}
\begin{minipage}[htb]{0.97\linewidth}
\centering\includegraphics[width=0.9\columnwidth]{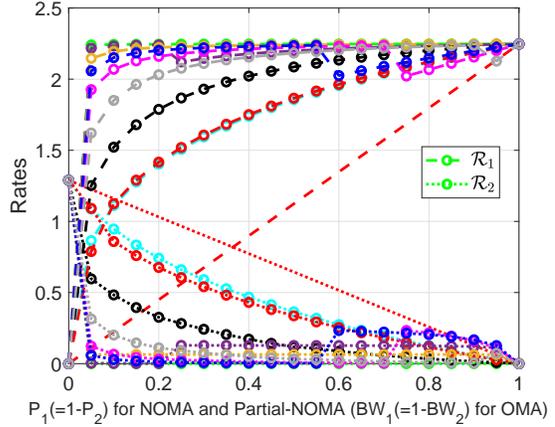}
\subcaption{Individual UE Throughput. Dashed (dotted) lines are used for $\mathcal{R}_1$ ($\mathcal{R}_2$).}\label{R1R2VsP}
\end{minipage}
\caption{Rates vs. $P_1(=1-P_2)$ for different $\alpha$ with corresponding optimum $\beta$, $\theta_1$, $\theta_2$ to maximize $\mathcal{R}_{\rm tot}$.}\label{RatesVsP}
\end{figure}

Fig. \ref{RatesVsP} is an abstraction of the rate region typically plotted for NOMA and OMA. For NOMA and partial-NOMA, the rate region plots the throughput of $\text{UE}_1$ against the throughput of $\text{UE}_2$ for every $P_1(=1-P_2)$ from 0 to 1. For OMA, on the other hand, the rates are plotted for every $\text{BW}_1 (=1-\text{BW}_2)$; of course OMA transmissions enjoy full power. {It should be noted that, although traditionally OMA is defined as one UE having access to the entire resource-block, in the context of rate regions for the sake of comparison, two UEs share non-overlapping fractions of the same resource-block. In our work, since partial-NOMA UEs have their overlap in the frequency domain, OMA UEs share non-overlapping fractions of the frequency channel, i.e., non-overlapping fractions of the bandwidth $\text{BW}_1$; thus, $\text{BW}_2=1-\text{BW}_1$. This way the rate region compares the performance of the following: 1) OMA: where two UEs split the bandwidth resource in a non-overlapping fashion while each utilizing the full power (i.e., $P=1$) for their transmissions, 2) NOMA: where two UEs fully share the bandwidth resource while splitting the power among themselves for the transmission of their messages, i.e., $P_1 \leq 1$, $P_2 \leq 1$ and $P_1+P_2=1$. Thus, in the literature, rate regions are used to show the superiority of NOMA due to the full spectrum reuse despite the lower transmit powers of the individual UE messages and introduction of intracell interference, over OMA where there are higher (full) transmit powers of individual UE messages and no intracell interference but no spectrum reuse. Here, we aim to shed light on the case of: 3) partial-NOMA: {where two UEs split the power among their messages like in the case of traditional NOMA; however, they share only an overlap $\alpha$ of the bandwidth, resulting in both spectrum reuse and intracell interference lower than traditional NOMA}. It should be noted that as there is no overlap of bandwidth in OMA, the OMA case may seem similar to partial-NOMA with $\alpha=0$. However, the two are different as each UE in OMA has messages with full power, i.e., $P=1$ while in partial-NOMA with $\alpha=0$, due to the definition of the setup which is for general $\alpha$, $P_1 \leq 1$, $P_2 \leq 1$ and $P_1+P_2=1$. Additionally, the rate region curves for OMA are plotted against $\text{BW}_1$, while the rate region curves for partial-NOMA (including $\alpha=0$), similar to those for traditional NOMA, are plotted against $P_1$.} 

Since the rate region for different values of $\alpha$ can be difficult to read in the partial-NOMA setup, we abstract the rate region by plotting the cell sum rate and the individual UE throughput against increasing $P_1$ for NOMA and partial-NOMA (against increasing $\text{BW}_1$ for OMA) in Figs. \ref{TotRateVsP} and \ref{R1R2VsP}, respectively. It ought to be mentioned that in addition to using the optimum $\theta_1$ and $\theta_2$ that maximize $\mathcal{R}_{\rm tot}$, as is typical for obtaining the boundary of the rate region, the partial-NOMA setup also involves using the optimum $\beta$ for each value of $\alpha$ and $P_1$. We consider BS intensity $\lambda=10$, noise power $\sigma^2=-90$ dB and $\eta=4$. 

From Fig. \ref{TotRateVsP}, we observe that {partial-NOMA with lower values of $\alpha$ outperforms traditional NOMA in terms of $\mathcal{R}_{tot}$.} It is, however, important to mention that partial-NOMA is able to do so because it has the advantage of modified-SIC decoding courtesy of the received filtering. We also observe that all values of $\alpha$ still outperform OMA significantly in terms of $\mathcal{R}_{\rm tot}$. {Additionally, by increasing $\alpha$ from 0, we observe that $\mathcal{R}_{\rm tot}$ first increases with $\alpha$, followed by a decrease in $\mathcal{R}_{\rm tot}$ with $\alpha$, and then an increase to $\alpha=1$. Note that the value of $\alpha$ until which $\mathcal{R}_{\rm tot}$ increases initially, grows with $P_1$. This trend of an increase in $\mathcal{R}_{\rm tot}$ with $\alpha$ at first followed by a decrease can be attributed to the trade off between spectrum reuse and interference. In the low $\alpha$ regime, increasing $\alpha$ does not increase $\mathcal{I}(\alpha,\beta)$ significantly and the impact from the resulting increase in interference is lower than the impact of the gains from the increased spectrum reuse with $\alpha$. This results in an increase in $\mathcal{R}_{\rm tot}$ with $\alpha$. After a certain $\alpha$, the impact of $\mathcal{I}(\alpha,\beta)$ becomes more significant and the impact of the increasing interference with $\alpha$ is more dominant than the impact of the increasing spectrum reuse\footnote{Note that FSIC plays an important role in this; without it, the impact of spectrum reuse would always be more significant.}. We thus observe a decrease in $\mathcal{R}_{\rm tot}$ with $\alpha$. At $\alpha=1$, although interference is maximum, the impact of full spectrum reuse between the two UEs is more significant, resulting in an increase in $\mathcal{R}_{\rm tot}$.} {These trends shed light on the existence of a range of smaller values of $\alpha$ which are superior to traditional NOMA in terms of $\mathcal{R}_{\rm tot}$ because of its spectrum reuse and interference trade off, followed by a range of larger $\alpha$ values that are inferior. This also highlights the importance of the careful choice of $\alpha$ required for different network goals.}

Fig. \ref{R1R2VsP} shows that in the case of OMA, the throughput of $\text{UE}_1$ is inferior to any partial-NOMA or NOMA setup while its throughput for $\text{UE}_2$ is superior to any partial-NOMA or NOMA. As $\alpha$ increases from 0 to 1, the throughput of $\text{UE}_1$ ($\text{UE}_2$) decreases (increases). This highlights the unexpected observation that, in terms of the individual UE throughput, traditional NOMA ($\alpha=1$) is closer to OMA than partial-NOMA with an overlap $\alpha<1$. Since the rate region reflects the boundaries of achievable throughput, these results show the ability of partial-NOMA to achieve more disparate performance than the other two schemes, highlighting the potential for greater flexibility. {Also, note that for values of $\alpha$ such as $\{0.1, 0.2, 0.25, 0.35\}$ $\mathcal{R}_1$ ($\mathcal{R}_2$) does not increase (decrease) monotonically with $P_1$. A significant change is seen at these values which corresponds to a switch in the decoding technique from $\bar{M}_1=M_1$ to $\bar{M}_1=M_0$ (i.e., from using traditional SIC to when $\text{UE}_1$ does not decode the message of $\text{UE}_2$). This is because at higher $P_1$, with these relatively smaller $\alpha$ values, decoding the message of the weak UE becomes inefficient so $\text{UE}_1$ starts treating the message of $\text{UE}_2$ as noise. Note that smaller (larger) $\alpha$ values have $\bar{M}_1=M_0$ ($\bar{M}_1=M_1$) for all $P_1$.} 


It is important to mention that Fig. \ref{RatesVsP} is plotted to maximize $\mathcal{R}_{\rm tot}$; in the case of traditional NOMA, $\beta$ can only take on the value 0 and so the rate region can be used to identify the maximum achievable throughput of a TMT constrained setup. However, when $\alpha<1$, the selected $\beta$ impacts performance and so the results in Fig. \ref{RatesVsP} cannot be used to see the gains that would be achievable from a TMT constrained setup.

\subsection{Problem Formulation $-$ Constrained Cell Sum Rate Maximization}
As the abstraction of the rate region plotted in the previous subsection aims to maximize the unconstrained cell sum rate for a given $\alpha$, in this subsection we formulate a problem for a more practical setup where a TMT is required to be achieved by each UE. Accordingly, we formally state the problem as follows:
\begin{itemize}
\item $\mathcal{P}1$ - Maximum cell sum rate, given $\alpha$, subject to the TMT $\mathcal{T}$:
\begin{align*}
\max\limits_{ (P_1, \theta_1,\theta_2,\beta)} &  \mathcal{R}_{\rm tot} \\
 \text{ subject to: } & \sum\limits_{i=1}^2 P_i= 1 \\
&   0 \leq \beta \leq \beta_{\rm max} \\
&  \mathcal{R}_i \geq \text{$\mathcal{T}$}, \; i \in \{1,2\},
\end{align*}
\end{itemize}
where $\beta_{\rm max}=1-\alpha$. It is evident that the constraints in $\mathcal{P}1$ are not affine, and therefore, the problem is non-convex. Thus, an optimal solution,
{i.e., choice of $\beta$, $P_1=(1-P_2)$ and $\theta_i$ for $i \in \{1,2\}$ that results in the maximum constrained $\mathcal{R}_{\rm tot}$,} can only be obtained by using an exhaustive search.

\subsection{Efficient Algorithm}\label{4c}
As has been mentioned, only an exhaustive search over all combinations of $P_1=(1-P_2)$, $\theta_1$, $\theta_2$ and $\beta$ can guarantee the optimum resource allocation for the above problem. In this subsection, we propose an algorithm based on intuition. While we cannot guarantee our algorithm to be optimum, it provides a feasible solution to meet the constraints of the problem. 


The following is known:
\begin{enumerate}
\item From the rate region for static channels in traditional NOMA, a resource allocation (RA) that results in the weak UE achieving the TMT $\mathcal{T}$, while all of the remaining power being allocated to the strong UE to maximize its throughput, is the optimum solution for that problem. An example of this is presented in \cite{N14}. Extending this to our large-scale partial-NOMA setup, for a given $\alpha$ and $\beta$, i.e., fixed bandwidths for the two UEs, the optimum RA would require achieving TMT for the weak UE and using the remaining power to maximize $\mathcal{R}_1$.
\item The impact of bandwidth is generally more significant on throughput than the impact of power, as throughput grows linearly with bandwidth but only as the logarithm with power.
\end{enumerate}

\textbf{\emph{Remark 3:}} In regard to 1), for traditional NOMA, achieving TMT for $\text{UE}_2$ and maximizing $\mathcal{R}_{\rm tot}$ would require allocating the smallest $P_2$ required to achieve $\mathcal{R}_2=\mathcal{T}$ \cite{myNOMA_icc,my_nomaMag} so that the largest $P_1$ possible would be left for $\text{UE}_1$ to maximize its throughput with. This is because a strong UE with its superior channel can obtain more from any power than the weak UE, and hence the least required power should be spent on the weak UE. However, in the partial-NOMA setup, the minimum $P_2$ that achieves TMT for the weak UE may result in $\text{UE}_1$ being in outage because of the impact of received filtering and corresponding $\mathcal{I}(\alpha, \beta)$. While we still want to allocate least $P_2$, it may need to be increased so that $\text{UE}_1$ can be in coverage as will be explained.

\textbf{\emph{Remark 4:}} In light of 2), it may be tempting to think that the largest $\beta$, i.e., $\beta_{\rm max}$ (corresponding to the smallest $\text{BW}_2$), with the smallest $P_2$ that can achieve TMT for the weak UE will result in the optimum solution. However, if $\alpha$ is small, the largest $\beta$ may result in insufficient bandwidth for $\text{UE}_2$. This may cause it to either not be able to meet TMT at all or to compensate for the small $\text{BW}_2$ by using a very large $P_2$ to achieve TMT. The latter would result in very little $P_1$ being left for $\text{UE}_1$ resulting in very low $\mathcal{R}_1$ because of low coverage despite having a large $\text{BW}_1$. 

Hence, we propose opting for an RA strategy that aims to find the lowest $P_2$ required to meet TMT for $\text{UE}_2$ and obtain the maximum $\mathcal{R}_1$ (and therefore $\mathcal{R}_{\rm tot}$) for each value of $\beta$, starting from $\beta_{\rm max}$ and decreasing it. Starting from the largest $\beta$, $\beta_{\rm max}$, this is done until $\mathcal{R}_{\rm tot}$ starts decreasing. At this point we have found the optimum $\beta$ because further decreasing $\beta$ will only deteriorate $\mathcal{R}_{\rm tot}$. The optimum RA is then selected by choosing the $\beta$ and its corresponding $P_1 (=1-P_2)$, $\theta_1$ and $\theta_2$ that result in the largest $\mathcal{R}_{\rm tot}$. 

{Using the definitions of $\mathcal{R}_1$ and $\mathcal{R}_2$ based on the developed analysis in Section III,} our algorithm for $\mathcal{P}1$ thus solves the problem
\begin{align*}
&\max\limits_{ (P_1, \theta_1,\theta_2,\beta)}   \mathcal{R}_1 \\
& \text{ subject to: }  \sum\limits_{i=1}^2 P_i= 1, \; 0 \leq \beta \leq 1-\alpha, \text{ and } \mathcal{R}_2=\mathcal{T}.
\end{align*}

Given $\alpha$ and $\beta$, we first search for the minimum $P_2$ that allows $\text{UE}_2$ to attain a throughput equal to the TMT; this leaves the largest possible $P_1$ for $\text{UE}_1$. Corresponding to this $P_2$, $\text{UE}_1$ can be in the following three states:
\begin{itemize}
\item State I ($\tilde{P}_2^1>0$): If $\tilde{P}_1 \leq 0$, increasing $P_2$ makes $\tilde{P}_1$ more negative and can therefore not impact $\mathcal{R}_1$. If $\tilde{P}_1>0$, increasing $P_2$ will make $\tilde{P}_1$ smaller and consequently $M_0$ larger which will not result in its selection for a potentially larger $\mathcal{R}_1$. Hence, if $\tilde{P}_2^1>0$ for the minimum $P_2$ that can achieve $\mathcal{R}_2=\mathcal{T}$, the optimum $P_2$ and $\theta_2$ have been found, the corresponding optimum $\theta_1$ that maximizes $\mathcal{R}_1$ should be selected to maximize $\mathcal{R}_{\rm tot}$.

\item State II ($\tilde{P}_2^1<0$ and $\tilde{P}_1>0$):
\begin{enumerate}
\item Optimize $\theta_1$ to maximize $\mathcal{R}_1$ and store as $\theta_{1,II}$ and $\mathcal{R}_{1,II}$, respectively. Note that here $M_0$ is selected as $\tilde{P}_2^1<0$ and so $M_1$ is not possible.
\item Increase $P_2$ until $\tilde{P}_2^1>0$ and $M_1$ is selected. Calculate the corresponding $\mathcal{R}_1$ using the optimum $\theta_1$ that maximizes it and store as $\mathcal{R}_{1,I}$ and $\theta_{1,I}$, respectively. If $M_1$ is never selected, $\mathcal{R}_{1,I}=0$.
\item Store the larger of the two throughputs $\mathcal{R}_{1,I}$ and $\mathcal{R}_{1,II}$, and its corresponding transmission rate as $\mathcal{R}_{1}$ and $\theta_{1}$, respectively.
\end{enumerate}

\item State III ($\tilde{P}_2^1<0, \tilde{P}_1 \leq 0$): As $\text{UE}_1$ is in outage in this state, increase $P_2$ until state I or II is achieved and follow the corresponding steps.
\end{itemize}
We formally state the working in \textbf{Algorithm 1}.

{\setstretch{1.0}
\begin{algorithm*}
\caption{RA for a feasible solution to $\mathcal{P}1$}
\begin{multicols}{2}
\begin{algorithmic}[1]
\STATE $\beta_{\rm max}=1-\alpha$, $\Delta_{\beta}=\beta_{\rm max}/10$, $\mathcal{R}_1^{\beta, \rm{vec}}=[\;]$, $\theta_1^{\beta, \rm{vec}}=[\;]$, $P_1^{\beta, \rm{vec}}=[\;]$, $\mathcal{R}_2^{\beta, \rm{vec}}=[\;]$, $\theta_2^{\beta, \rm{vec}}=[\;]$
\FOR{$\beta=\beta_{\rm max}:-\Delta_{\beta}:0$} \label{stepK3}
\STATE $\rm{flag}_1=0$, State=[ ]
\FOR{{$P_2=0:\Delta_P:1$}} \label{stepK2}
\FOR{ {$\theta_2=\theta_{LB}:\Delta_{\theta}:\theta_{UB}$}}
\STATE Calculate $\mathcal{R}_2$ using \eqref{R_i} with \eqref{cvg} and \eqref{Mbar_2}
\IF{ $\mathcal{R}_2 \geq \mathcal{T}$} 
\IF{ $\tilde{P}_2^1>0$} 
\STATE State=I
\ENDIF
\STATE Go to \ref{stepK1}
\ENDIF
\ENDFOR
\STATE $\rm{flag}_2=0$ 
\IF{ $\mathcal{R}_2 < \mathcal{T}$} 
\IF{$P_2=1$}
\STATE $\rm{flag}_2=1$ 
\ELSE
\STATE Go to \ref{stepK2}
\ENDIF
\ENDIF
\IF{$\rm{flag}_2=0$} \label{stepK1}
\IF{State=I}
\FOR{$\theta_1=\theta_{LB}:\Delta_{\theta}:\theta_{UB}$}
\STATE Calculate $\mathcal{R}_1$ using \eqref{R_i} with \eqref{cvg} and \eqref{Mbar_1}
\STATE Update $\mathcal{R}_{1,I}^{\rm vec}=[\mathcal{R}_{1,I}^{\rm vec}; \mathcal{R}_1]$
\ENDFOR
\STATE Store $\mathcal{R}_{1,I}=\max (\mathcal{R}_{1,I}^{\rm vec})$ and corresponding $\theta_{1,I}$, $P_{1,I}$, $\mathcal{R}_{2,I}$, $\theta_{2,I}$
\ELSE
\FOR{$\theta_1=\theta_{LB}:\Delta_{\theta}:\theta_{UB}$}
\IF{ $\tilde{P}_1>0$}
\STATE State=II
\IF{$\rm{flag}_1=0$}
\STATE Calculate $\mathcal{R}_1$ using \eqref{R_i} with \eqref{cvg} and \eqref{Mbar_1}
\STATE Update $\mathcal{R}_{1,II}^{\rm vec}=[\mathcal{R}_{1,II}^{\rm vec}; \mathcal{R}_1]$
\ENDIF
\ENDIF
\ENDFOR
\STATE Store $\mathcal{R}_{1,II}=\max (\mathcal{R}_{1,II}^{\rm vec})$ and corresponding $\theta_{1,II}$, $P_{1,II}$, $\mathcal{R}_{2,II}$, $\theta_{2,II}$
\IF{$\mathcal{R}_{1,II}=0$}
\STATE State=III 
\STATE Go to \ref{stepK2}
\ELSE
\STATE $\rm{flag}_1=1$
\IF{$P_2<1$}
\STATE Go to \ref{stepK2}
\ENDIF
\ENDIF
\ENDIF
\ENDIF
\IF{$\rm{flag}_1=1$}
\STATE Store $\mathcal{R}_1^{\beta, \rm{vec}}=[\max(\mathcal{R}_{1,I},\mathcal{R}_{1,II}); \mathcal{R}_1^{\beta, \rm{vec}} ]$ and corresponding $\theta_1^{\beta, \rm{vec}}$, $P_1^{\beta, \rm{vec}}$, $\mathcal{R}_2^{\beta, \rm{vec}}$, $\theta_2^{\beta, \rm{vec}}$
\STATE Go to \ref{stepK5}
\ELSE
\IF{$\rm{flag}_2=1$}
\STATE TMT cannot be met by $\text{UE}_2$
\STATE Go to \ref{stepK3}\label{stepK4}
\ELSE
\STATE Store $\mathcal{R}_1^{\beta, \rm{vec}}=[\mathcal{R}_{1,I}; \mathcal{R}_1^{\beta, \rm{vec}} ]$ and corresponding $\theta_1^{\beta, \rm{vec}}$, $P_1^{\beta, \rm{vec}}$, $\mathcal{R}_2^{\beta, \rm{vec}}$, $\theta_2^{\beta, \rm{vec}}$
\STATE Go to \ref{stepK5}
\ENDIF
\ENDIF
\ENDFOR
\IF{$\beta>\beta_{\rm max}$}\label{stepK5}
\IF{$\mathcal{R}_1^{\beta, \rm{vec}}(\rm{end})<\mathcal{R}_1^{\beta, \rm{vec}}(\rm{end-1})$}
\STATE Store $\mathcal{R}_1=\max(\mathcal{R}_1^{\beta, \rm{vec}})$ and corresponding $\theta_1$, $P_1$, $\mathcal{R}_2$, $\theta_2$, $\beta$.
\IF{$\mathcal{R}_1<\mathcal{T}$}
\STATE TMT cannot be met by $\text{UE}_1$
\ENDIF
\STATE $\beta$ that maximizes $\mathcal{R}_{\rm tot}$ found; \textbf{exit}\label{stepK6}
\ENDIF
\ENDIF
\ENDFOR
\end{algorithmic}\label{algo1}
\end{multicols}
\end{algorithm*}
}
In \textbf{Algorithm \ref{algo1}}, $\rm{flag}_1=0$ denotes that State II has not been achieved as of yet for the current value of $\beta$ while $\rm{flag}_1=1$ denotes that State II has been achieved at least once. If the power budget has been expended but $\text{UE}_2$ cannot meet the TMT, $\rm{flag}_2=1$; otherwise, $\rm{flag}_2=0$. Thus if $\rm{flag}_2=1$, $\beta$ needs to be decreased to give $\text{UE}_2$ a larger bandwidth to achieve TMT; if $\beta$ is decreased to 0 but $\text{UE}_2$ can still not achieve TMT, the TMT is too high to be met by the system and ought to be decreased. Since the range of possible $\beta$ changes with $\alpha$, we standardize $\Delta_{\beta}$ to be a function of $\beta_{\rm max}$ in line 1 so that we select from a fixed number of $\beta$ values irrespective of $\alpha$ for fairness between different values of $\alpha$. Similarly, since the range of transmission rates is $\theta_i \geq 0$ $i \in \{1,2\}$, we make our search finite by searching in the range $\theta_{LB} \leq \theta_i \leq \theta_{UB}$, increasing in steps of $\Delta_{\theta}$; $P_2$ is also increased in steps of $\Delta_P$.

Given $\alpha$ and $\mathcal{T}$, \textbf{Algorithm 1} starts with $\beta_{\rm max}$, the largest value of $\beta$, in line \ref{stepK3}. For a $\beta$, it searches for the smallest $P_2$ and the corresponding lowest $\theta_2$ that can attain the TMT. If $\text{UE}_1$ can be in State I with the selected $P_2$ and $\theta_2$, $\theta_1$ is optimized to maximize $\mathcal{R}_1$ and the optimum parameters that maximize $\mathcal{R}_{\rm tot}$ have been found for this $\beta$. However, if State I is not achieved, but State II is achieved, we optimize $\theta_1$ to maximize $\mathcal{R}_1$ and store it as $\mathcal{R}_{1,II}$. $P_2$ is then increased until State I can be achieved. If it is achieved before exhausting the power budget, the corresponding $\theta_1$ is optimized to maximize $\mathcal{R}_1$ and stored as $\mathcal{R}_{1,I}$; $\mathcal{R}_{1,I}$ and $\mathcal{R}_{1,II}$ are compared to see which is larger and the corresponding parameters $P_1=(1-P_2)$, $\theta_1$ and $\theta_2$ are stored as the optimum for this $\beta$. If State I cannot be achieved, $\mathcal{R}_{1,II}$ and its corresponding parameters are stored. If we are in State III, $P_2$ is increased until State I or II is achieved and the corresponding steps are followed. However, if the TMT cannot be met for $\text{UE}_2$ using full power, i.e., we are in State III even when $P_2$ is increased to 1 in line \ref{stepK4}, $\text{BW}_2$ is not sufficient, $\rm{flag}_2=1$, and we go to line \ref{stepK3} to reduce $\beta$. If the TMT is met by $\text{UE}_2$ and $\mathcal{R}_1^{\beta, \rm{vec}}$ is stored for the iteration of $\beta$, we go to line \ref{stepK5}. If this is the first iteration of the $\beta$ loop, we go to the next iteration. If it is not the first iteration and if the throughput of $\text{UE}_1$ calculated is larger than that in the last iteration of the $\beta$ loop, we again go to the next iteration of the $\beta$ loop. However, if it is not the first iteration of the $\beta$ loop, and the throughput of $\text{UE}_1$ calculated is smaller than that in the last iteration, the optimum throughput of $\text{UE}_1$ has already been found and we store it as $\mathcal{R}_1$ along with its corresponding $\mathcal{R}_2$ and parameters $P_1=(1-P_2)$, $\theta_1$, $\theta_2$, and $\beta$. We check to see if $\text{UE}_1$ is able to meet the TMT; however, it should be noted that increasing $\beta$ will not improve the throughput of $\text{UE}_1$ further irrespective of whether the TMT has been met or not. We thus exit the algorithm in line \ref{stepK6}.

{Since the purpose of \textbf{Algorithm 1} is to provide an efficient alternative to an exhaustive search in terms of complexity, it is important to define a measure of complexity to compare the two. We measure complexity in terms of the sum of the number of times a UE's throughput $\mathcal{R}_i$, $i \in \{1,2\}$, is calculated. The algorithm iterates over the number of $\beta$, power, and transmission rate combinations. As has been mentioned, our algorithm searches in $\theta_{LB} \leq \theta_i \leq \theta_{UB}$ in steps of $\Delta_{\theta}$ to make the search finite. Similarly, the algorithm searches for $0\leq P_2 \leq 1$ in steps of $\Delta_P$. When $\alpha<1$, the algorithm searches for $\beta_{\rm max} \leq \beta \leq 0$ in steps of $\Delta_{\beta}=\beta_{\rm max}/10$; for $\alpha=1$, there is only one choice of $\beta$ which is 0. For a fair comparison, we use the same search space for the exhaustive search. In particular, there are $\hat{\Delta}_{\theta}=(\theta_{UB} - \theta_{LB})/\Delta_{\theta} +1$ choices of $\theta_i$, $i \in \{1,2\}$ and $\hat{\Delta}_P=1/\Delta_P +1$ choices of $P_2$. When $\alpha=1$, there is one choice of $\beta$ and so $\hat{\Delta}_{\beta}=1$. However, when $\alpha<1$, there are $\hat{\Delta}_{\beta}=\beta_{\rm \max}/\Delta_{\beta} +1$ choices of $\beta$. Note that since we have fixed $\Delta_{\beta}$ to $\beta_{\rm max}/10$, $\hat{\Delta}_{\beta}=11$ when $\alpha<1$. While \textbf{Algorithm 1} does not go through all combinations of these choices, an exhaustive search does. Hence, the complexity of an exhaustive search for this setup is $ \hat{\Delta}_{\beta} \hat{\Delta}_P \hat{\Delta}_{\theta}^2$.}

\noindent
\textbf{\em Remark 5}: {Each value of $\alpha$ requires an exhaustive search over all combinations of $\beta$, $P_1=(1-P_2)$, $\theta_1$ and $\theta_2$ to obtain the contour plots against $\beta$ and $P_1$ that maximize $\mathcal{R}_{\rm tot}$. It is thus not possible to obtain these plots from an exhaustive search for many $\alpha$ values. However, we would like to mention that the results obtained from \textbf{Algorithm 1} matched those of the exhaustive search for the $\alpha$ values that we did conduct them for. While we can still not guarantee that our algorithm finds the optimum solution for all values of $\alpha$ and the TMT constraint, this highlights the accuracy of the feasible solution found by our algorithm.} 

\section{Results}
We consider BS intensity $\lambda=10$, noise power $\sigma^2=-90$ dB and $\eta=4$. In the results in Section \ref{ResultsA}, resource allocation is fixed and unless stated otherwise, we transmit using $P_1=1-P_2=1/3$ and use identical transmission rates for clarity of presentation. The SINR thresholds ($\theta_1$ and $\theta_2$) corresponding to the transmission rates are thus represented using $\theta$. The results in Section \ref{ResultsB} use resource allocation obtained from \textbf{Algorithm 1} for solving the optimization problem $\mathcal{P}1$.

\subsection{Fixed Resource Allocation}\label{ResultsA}
\begin{figure}
\begin{minipage}[htb]{0.97\linewidth}
\centering\includegraphics[width=0.9\columnwidth]{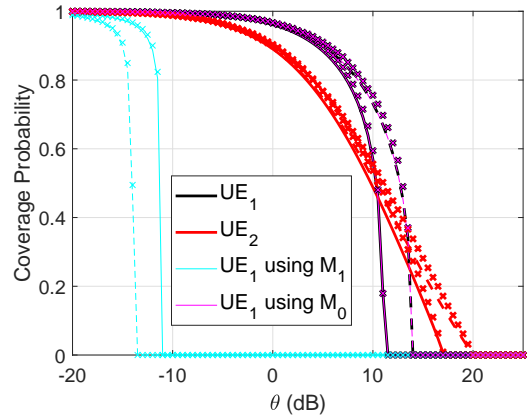}
\subcaption{$\alpha=0.25$}\label{alphaPoint25}
\end{minipage}
\begin{minipage}[htb]{0.97\linewidth}
\centering\includegraphics[width=0.9\columnwidth]{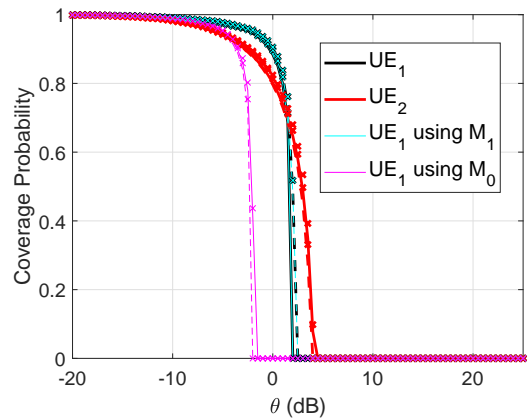}
\subcaption{$\alpha=0.75$}\label{alphaPoint75}
\end{minipage}
\caption{Coverage probabilities vs. $\theta$. Solid (dashed) lines represent $\beta=0$ ($\beta=(1-\alpha)/2$). Markers represent Monte Carlo simulations.}\label{ModelVal}
\end{figure}
Fig. \ref{ModelVal} is a plot of coverage probabilities against SINR threshold. Fig. \ref{alphaPoint25} uses $\alpha=0.25$ while Fig. \ref{alphaPoint75} uses $\alpha=0.75$; each plots the probabilities for $\beta=0$ and $\beta=(1-\alpha)/2$. The figure validates our analysis by using Monte Carlo simulations to show that the approximation in \textbf{Theorem 1} is tight. In addition to the coverage probabilities of the two UEs, the figure also plots the coverage probability of $\text{UE}_1$ using traditional SIC decoding (cyan curves), i.e., $\bar{M}_1=M_1$, and when $\text{UE}_1$ does not decode the message of $\text{UE}_2$ and treats the intracell interference from $\text{UE}_2$ as noise (magenta curves), i.e., $\bar{M}_1=M_0$. Since identical transmission rates are used for both UEs in this figure, the coverage of $\text{UE}_1$ for a given $\alpha$ is one or the other; however, if this was not the case, the coverage could have been equal to different decoding techniques at different $\theta$ depending on the selected (superior) technique in that case. 

{We observe that when $\alpha$ is small in Fig. \ref{alphaPoint25}, increasing $\beta$ from 0 to $(1-\alpha)/2$ increases the coverage probability as $\mathcal{I}(\alpha,\beta)$ decreases from 0 to $\beta_{\rm max}/2$ ($=(1-\alpha)/2$) as shown in Fig. \ref{intrfFactorVsBeta}.} {With the larger $\alpha$ used in Fig. \ref{alphaPoint75}, $\mathcal{I}(\alpha,\beta)$ increases from 0 to $\beta_{\rm max}/2$. Corresponding to this increase in interference, the coverage probability of $\text{UE}_2$, which treats the message of $\text{UE}_1$ as noise, decreases with $\beta$ from 0 to $\beta_{\rm max}/2$. $\text{UE}_1$, on the other hand, decodes the message of $\text{UE}_2$ as $\bar{M}_1=M_1$ in the case of Fig. \ref{alphaPoint75}; thus, as $\mathcal{I}(\alpha,\beta)$ increases from 0 to $\beta_{\rm max}/2$ decoding the message of $\text{UE}_2$ becomes easier for $\text{UE}_1$ and the coverage of $\text{UE}_1$ improves. Note that the case of $\text{UE}_1$ using $M_0$ follows a similar trend to $\text{UE}_2$ as it treats the message of the other UE as noise.} It should also be noted that the coverage probability for the UEs given an $\alpha$ is the same when $\beta=0$ and when $\beta=(1-\alpha)$. This is due to the symmetry of $\mathcal{I}(\alpha,\beta)$ about $\beta_{\rm max}/2$ which results in identical coverage for $\beta$ values of the form $(1-\alpha)x$ and $(1-\alpha)(1-x)$, where $x \in [0,1]$, due to identical values of $\mathcal{I}(\alpha,\beta)$ for such values of $\beta$. It is important to note that the throughput of the UEs will not be the same for such pairs of $\beta$ values. This is because, the $\beta$ value directly impacts bandwidth and therefore throughput; this will be observed in Figs. \ref{RtotVsAlpha} and \ref{RatesVsAlphaTh0} where different rates are observed for $\beta=0$ and $\beta=1-\alpha$ which have identical coverage. } 

\begin{figure}
\begin{minipage}[htb]{0.97\linewidth}
\centering\includegraphics[width=0.9\columnwidth]{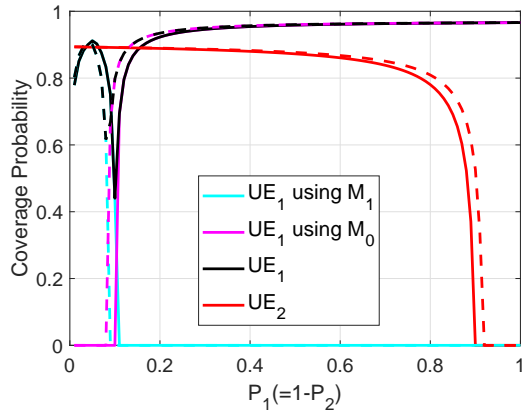}
\subcaption{{$\alpha=0.35$}}\label{vsP1alphaPoint35}
\end{minipage}
\begin{minipage}[htb]{0.97\linewidth}
\centering\includegraphics[width=0.9\columnwidth]{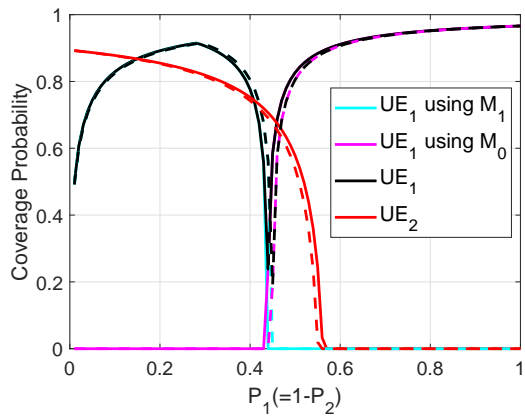}
\subcaption{$\alpha=0.75$}\label{vsP1alphaPoint75}
\end{minipage}
\caption{Coverage probabilities vs. $P_1(=1-P_
2)$ using $\theta=0$ dB. Solid (dashed) lines represent $\beta=0$ ($\beta=(1-\alpha)/2$).}\label{vsP1}
\end{figure}

{Fig. \ref{vsP1} plots the coverage probability of the UEs with increasing $P_1(=1-P_2)$ using $\theta=0$ dB and different $\beta$ values for {$\alpha=0.35$} and $\alpha=0.75$. As anticipated, the coverage of $\text{UE}_2$ decreases with $P_1$ because of deteriorating SINR for $\text{UE}_2$ as $P_1$ increases (i.e., $P_2$ decreases). Additionally, the coverage of $\text{UE}_2$ decreases with $\alpha$ due to the higher interference encountered as $\mathcal{I}(\alpha,\beta)$ increases with $\alpha$. In Fig. \ref{vsP1alphaPoint35}, we observe that the coverage of $\text{UE}_2$ increases with $\beta$, while in Fig. \ref{vsP1alphaPoint75}, we observe a slight decrease in coverage with $\beta$. This occurs because for lower (higher) values of $\alpha$, $\mathcal{I}(\alpha,\beta)$ decreases (increases) from $\beta=0$ to $\beta=(1-\alpha)/2$, thereby decreasing (increasing) interference. The coverage for $\text{UE}_1$ is more complex as the coverage first increases at low $P_1$ as the message of $\text{UE}_2$ is easily decoded due to high $P_2$ and then decreases as increasing $P_1$ makes decoding the message of $\text{UE}_2$ hard; $\bar{M}_1=M_1$ in this regime. After this, we observe a sharp increase in the coverage of $\text{UE}_1$ as the message of $\text{UE}_2$ is treated as noise with growing $P_1$ (and therefore, decreasing $P_2$) since $\bar{M}_1=M_0$. Note that in the region where $\text{UE}_2$'s message is being decoded by $\text{UE}_1$ (i.e., $\bar{M}_1=M_1$), when {$\alpha=0.35$}, having small $\mathcal{I}(\alpha, \beta)$ is a disadvantage as it reduces the power of the message of $\text{UE}_2$ being decoded; hence, we observe that $\beta=0$ outperforms $\beta=(1-\alpha)/2$ in this region. At a higher $P_1$, where the message of $\text{UE}_2$ is treated as noise, $\beta=(1-\alpha)/2$, with the smaller $\mathcal{I}(\alpha,\beta)$, outperforms $\beta=0$ as it experiences lower interference. The opposite trends hold in Fig. \ref{vsP1alphaPoint75} with the higher $\alpha$ value of 0.75; this is because here $\beta=(1-\alpha)/2$ has a higher interference factor $\mathcal{I}(\alpha,\beta)$ than $\beta=0$. }

\begin{figure}
\begin{minipage}[htb]{0.97\linewidth}
\centering\includegraphics[width=0.9\columnwidth]{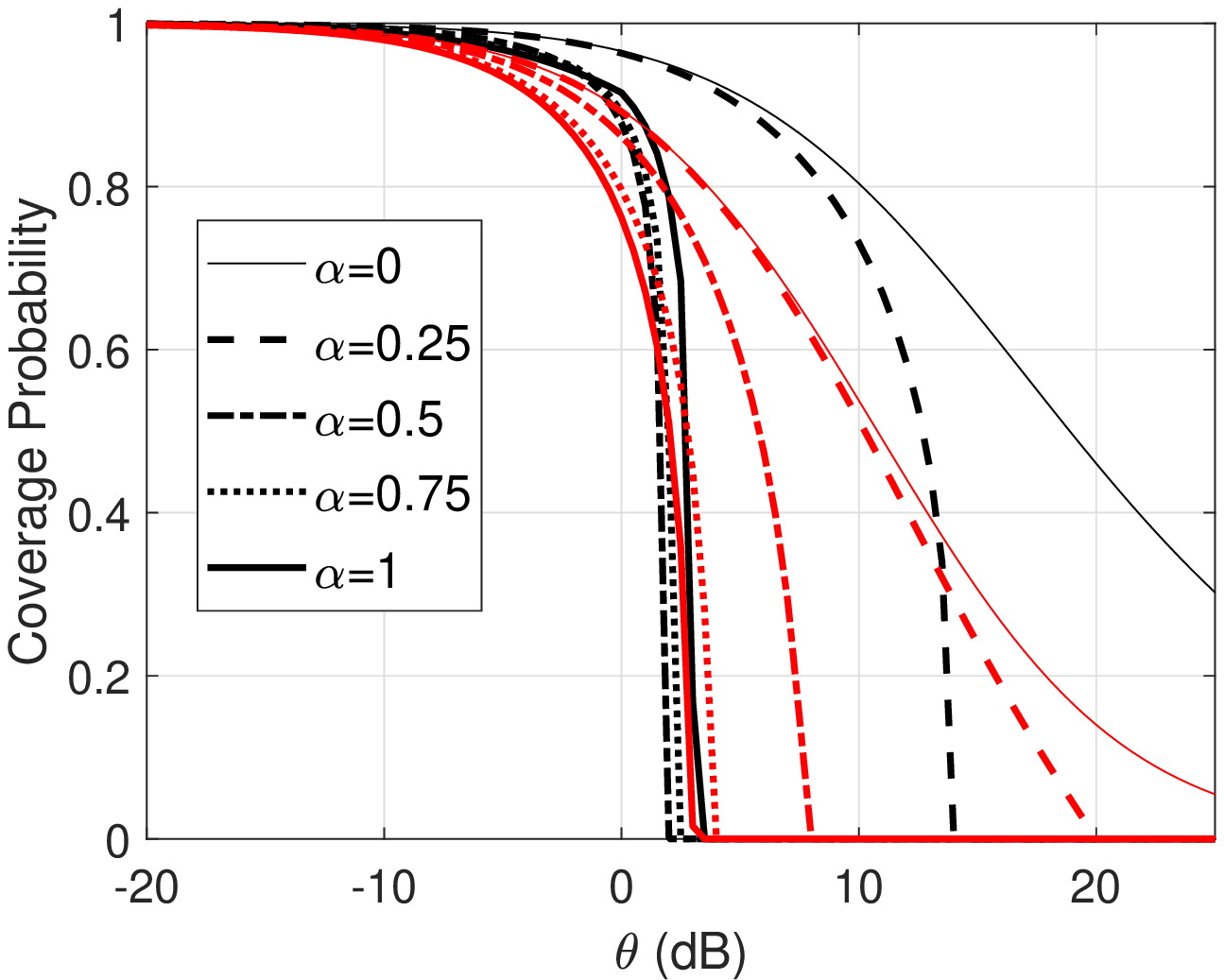}
\caption{Coverage probabilities vs. $\theta$ using {$\beta=(1-\alpha)/2$}. Black (red) lines represent $\text{UE}_1$ ($\text{UE}_2$).}\label{cvgVsThAllAlpha}
\end{minipage}\;\;\;\;
\begin{minipage}[htb]{0.97\linewidth}
\centering\includegraphics[width=0.9\columnwidth]{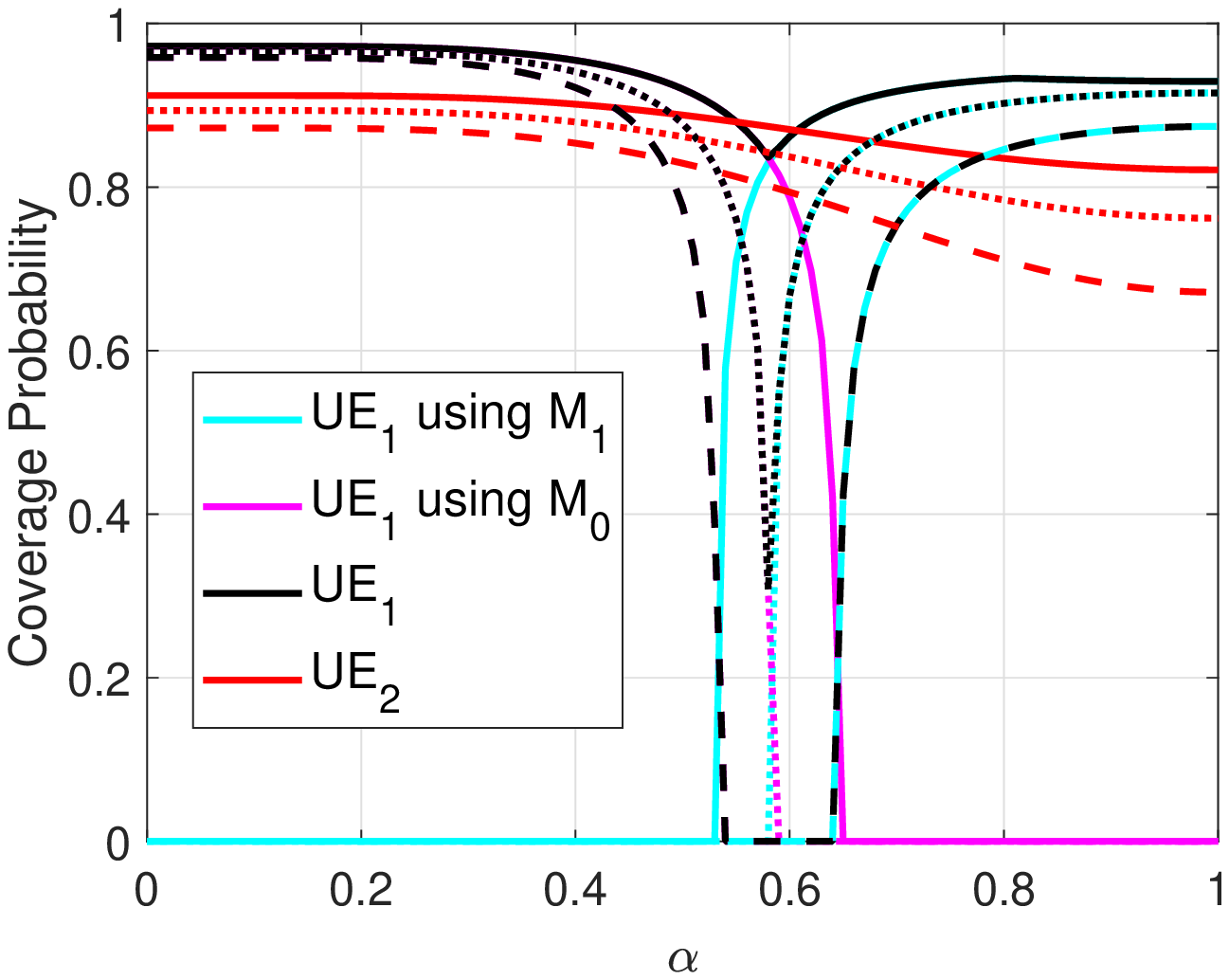}
\caption{Coverage probabilities vs. $\alpha$ using {$\beta=(1-\alpha)/2$}. Solid lines represent $\theta\!=\!-1$ dB, dotted represent $\theta\!=\!0$ dB and dashed represent $\theta\!=\!1$ dB.}\label{cvgVsAlphaAllTh}
\end{minipage}
\end{figure}
Fig. \ref{cvgVsThAllAlpha} is a plot of the coverage probabilities of the UEs vs. $\theta$ using {$\beta=(1-\alpha)/2$} for different values of $\alpha$. We observe that the coverage of $\text{UE}_2$ for any $\theta$ decreases as $\alpha$ increases. This is anticipated because $\mathcal{I}(\alpha, \beta)$ increases with $\alpha$; consequently, both intracell and intercell interference increase with $\alpha$ thereby reducing coverage. A different trend is observed for $\text{UE}_1$, on the other hand, where the coverage does not decrease monotonously with $\alpha$.

Fig. \ref{cvgVsAlphaAllTh}, a plot of coverage probability vs. $\alpha$, explains the above phenomenon better. Different $\theta$ values and {$\beta=(1-\alpha)/2$} are used. As before, the coverage probability of $\text{UE}_2$ decreases monotonically with $\alpha$. For $\text{UE}_1$, however, because of the employed modified-SIC decoding, there is a switch between not decoding the message of $\text{UE}_2$ (the curves corresponding to using $\bar{M}_1=M_0$) and employing traditional SIC decoding (the curves corresponding to $\bar{M}_1=M_1$). This results in a non-monotonic decrease in the coverage as $\alpha$ increases because the impact of increasing $\mathcal{I}(\alpha, \beta)$ is not as trivial as in the case of $\text{UE}_2$. We also observe that for larger values of $\theta$ (see $\theta=1$ dB), for some choices of power and transmission rate allocation, certain values of $\alpha$ will result in guaranteed outage {($0.54 \leq \alpha \leq 0.64$ for $\theta=1$ dB)}. This highlights the importance of careful resource allocation as well as parameter selection in a partial-NOMA setup to avoid guaranteed outage. 

It should also be mentioned that with appropriate resource allocation, partial-NOMA can result in better coverage than traditional NOMA ($\alpha=1$) for both UEs. Additionally, the curves for $\text{UE}_1$ using $M_1$ highlight the traditional SIC decoding scheme's inadequacy in the low $\alpha$ regime where, due to small $\mathcal{I}(\alpha, \beta)$, decoding the message of the weak UE becomes the bottleneck for coverage. Similarly, the curves for $\text{UE}_1$ using $M_0$ show that for higher values of $\alpha$ the intracell interference becomes significant and treating it as noise results in outage. 


Fig. \ref{RtotVsTheta} is a plot of the cell sum rate against $\theta$ using {$\beta=(1-\alpha)/2$}. The figure highlights that reducing $\alpha$ from the traditional NOMA setup ($\alpha=1$) increases the tolerance of the system to outage for high transmission rates. While traditional NOMA cannot support UEs that have messages with high transmission rates, instead of opting for such UEs to be designated an entire resource-block for the transmission of their messages, a more efficient approach is to use partial-NOMA where multiple UEs still share a resource-block and can transmit a message with high transmission rate. Essentially, the partial-NOMA setup is less restrictive in terms of the transmission rates that can be supported. {We also observe that the peak $\mathcal{R}_\mathrm{tot}$ first increases from $\alpha=0$, followed by a decrease and then an increase again to $\alpha=1$. Additionally, the peaks of the lower $\alpha$ values outperform that of traditional NOMA. This trend again highlights the existence of a range of $\alpha$ values which provide superior performance compared to traditional NOMA in terms of $\mathcal{R}_\mathrm{tot}$ followed by another range inferior to it.}

\begin{figure}
\begin{minipage}[htb]{0.97\linewidth}
\centering\includegraphics[width=0.9\columnwidth]{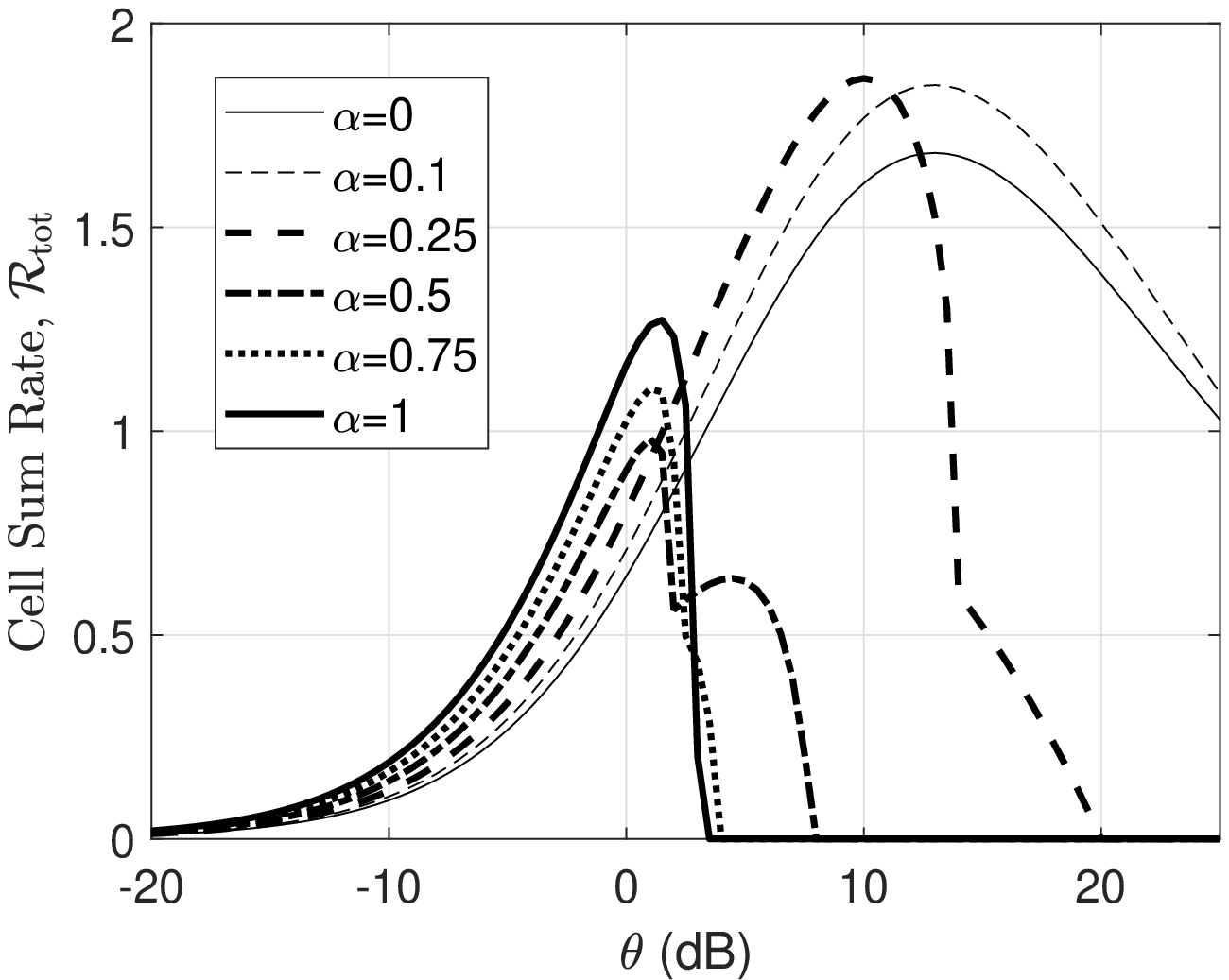}
\caption{Cell sum rate vs. $\theta$ using {$\beta=(1-\alpha)/2$}.}\label{RtotVsTheta}
\end{minipage}\;\;
\begin{minipage}[htb]{0.97\linewidth}
\centering\includegraphics[width=0.9\columnwidth]{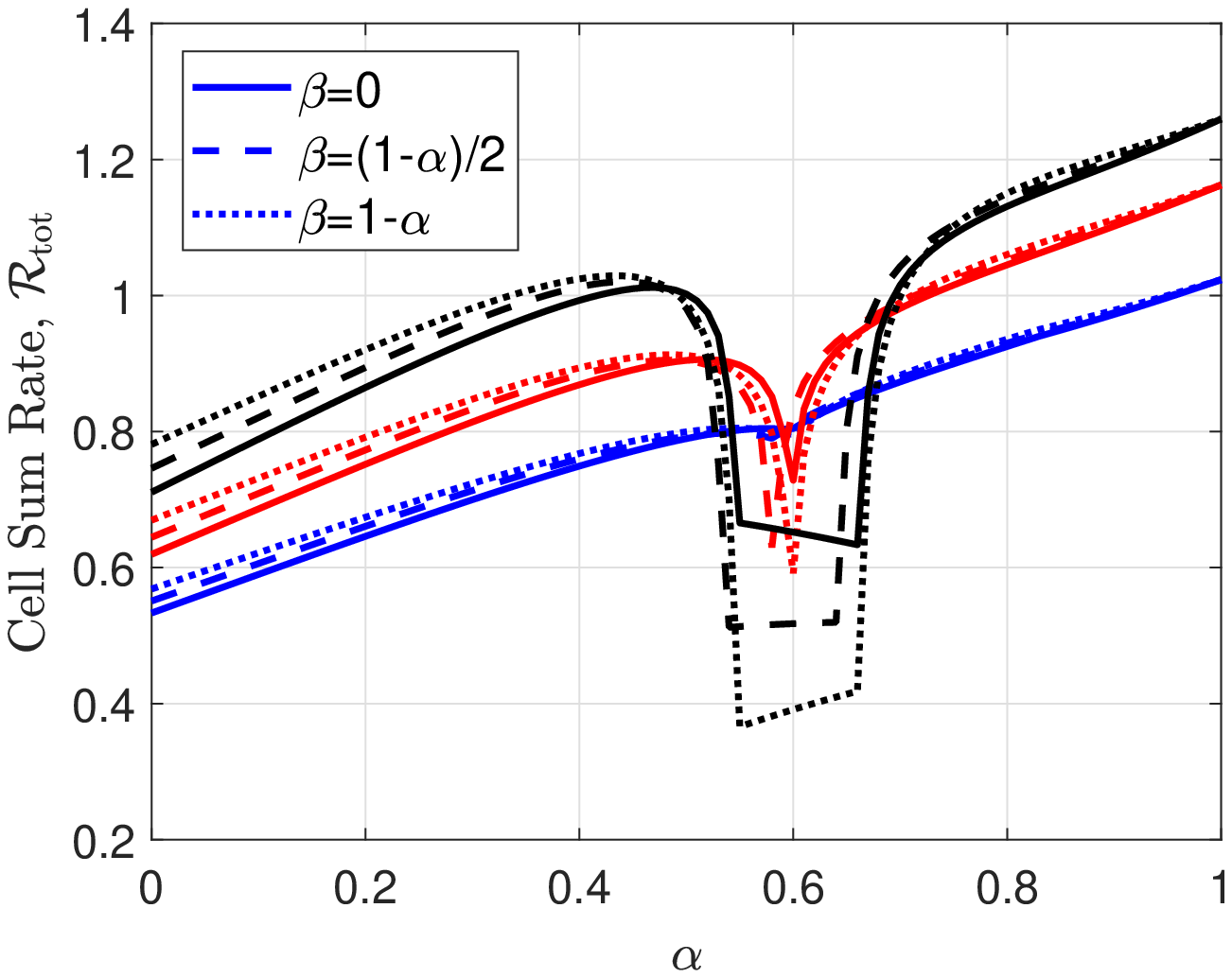}
\caption{Cell sum rate vs. $\alpha$. Blue lines represent $\theta=-1$ dB, red represent $\theta=0$ dB, and black represent $\theta=1$ dB.}\label{RtotVsAlpha}
\end{minipage}\;\;
\begin{minipage}[htb]{0.97\linewidth}
\centering\includegraphics[width=0.9\columnwidth]{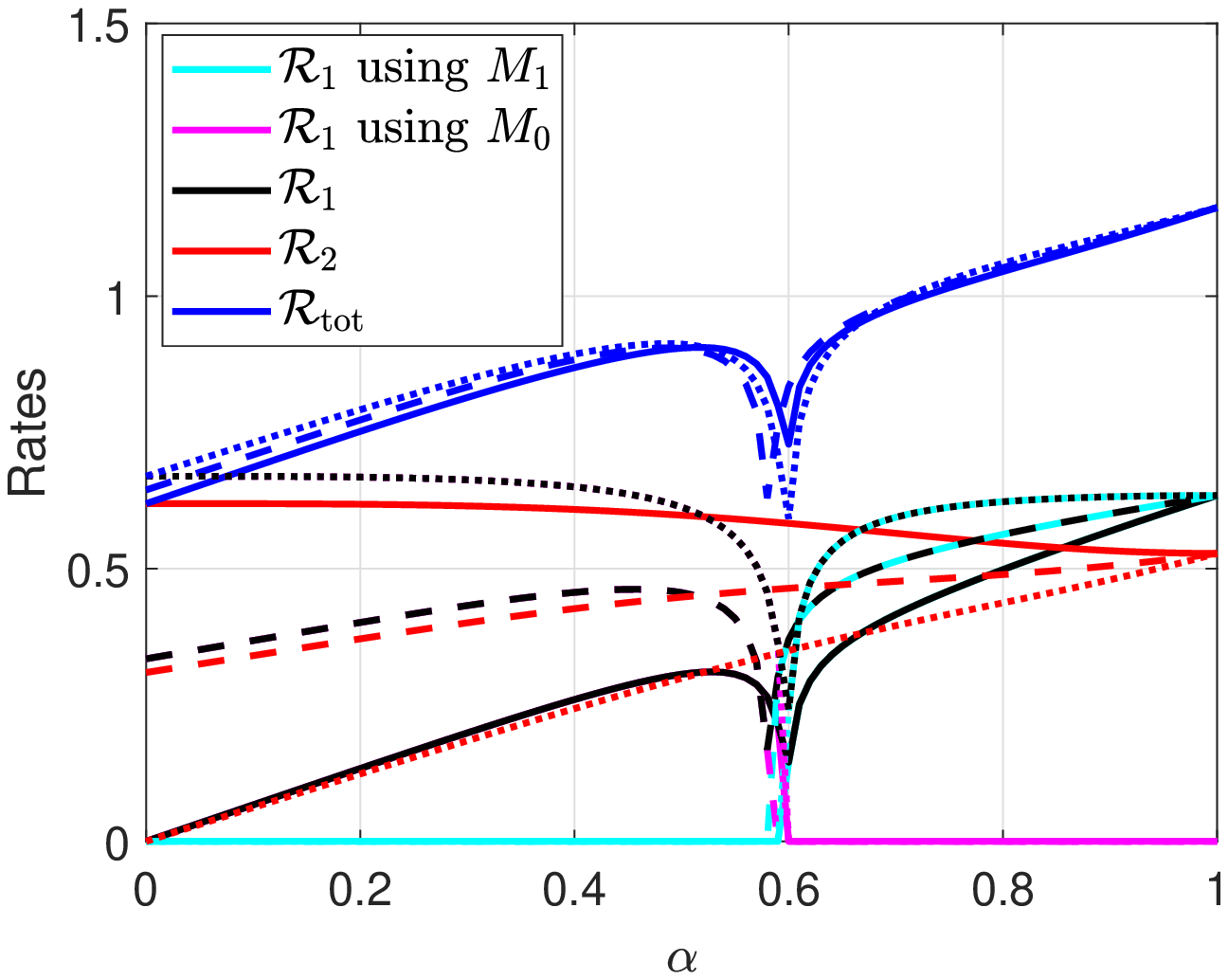}
\caption{Rates vs. $\alpha$ using $\theta=0$ dB. Solid lines represent $\beta=0$, dashed represent $\beta=(1-\alpha)/2$, and dotted represent $\beta=1-\alpha$.}\label{RatesVsAlphaTh0}
\end{minipage}
\end{figure}

Fig. \ref{RtotVsAlpha} is a plot of the cell sum rate against $\alpha$. Corresponding to the outage regions for $\text{UE}_1$ in Fig. \ref{cvgVsAlphaAllTh}, we observe dips in the cell sum rate. It is interesting to observe that $\theta$ values that support larger rates overall, such as $\theta=1$ dB have larger dips than lower $\theta$. This occurs because while the transmission rates being used make both $M_1$ and $M_0$ result in superior coverage conditions, since they are identical, they put a larger gap between the two conditions resulting in a larger region of outage for $\text{UE}_1$. This gap, and the consequent dip in rate, reduces as $\theta$ increases but the price paid is lower overall rate. Other than the dip caused by the outage region of $\text{UE}_1$, we observe that cell sum rate increases roughly linearly with $\alpha$. {Fig. \ref{RatesVsAlphaTh0} is plotted to gain better insight of why this occurs}

Fig. \ref{RatesVsAlphaTh0} is a plot of rates with increasing $\alpha$ using $\theta=0$ dB. For a given $\alpha$, the throughput of $\text{UE}_2$ ($\text{UE}_1$) decreases (increases) as $\beta$ increases because its bandwidth decreases (increases). For $\beta=0$ we observe that $\text{UE}_2$'s throughput decreases with $\alpha$. This is because, it has a fixed bandwidth of 1 in this case and its throughput is only impacted by coverage which decreases as $\mathcal{I}(\alpha, \beta)$ increases with $\alpha$. For the other two $\beta$ values, the throughput of $\text{UE}_2$ increases with $\alpha$ as the impact of the increasing bandwidth is greater than the increased interference resulting from higher $\mathcal{I}(\alpha, \beta)$. For the lower $\beta$ values, we observe that the throughput of $\text{UE}_1$ increases with $\alpha$ (other than the dips occurring from the outage region) as the impact of increasing bandwidth is larger than the increasing interference. When $\beta=1-\alpha$, however, the bandwidth is 1 and the throughput decreases with $\alpha$ because of the impact of increased interference occurring from $\mathcal{I}(\alpha, \beta)$. It should be noted that the rate of this decrease is much lower than the rate of increase from bandwidth gains for the other curves which is anticipated. This explains why an optimum $\alpha$, which is not 1, that maximizes cell sum rate is not observed in Fig. \ref{RtotVsAlpha} or \ref{RatesVsAlphaTh0}.


\subsection{Resource Allocation Using Algorithm 1}\label{ResultsB}

\begin{figure}
\begin{minipage}[htb]{0.97\linewidth}
\centering\includegraphics[width=0.83\columnwidth]{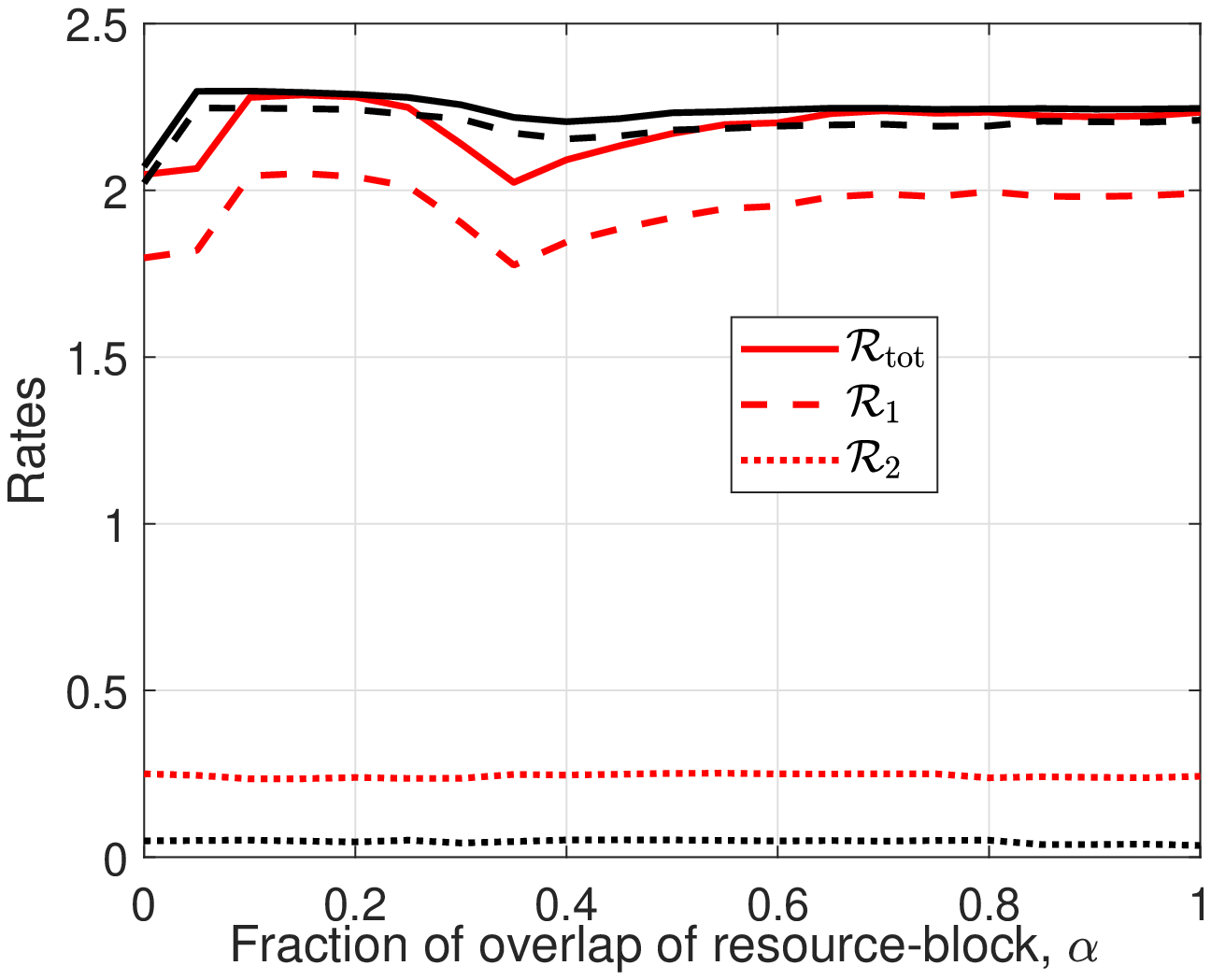}
\subcaption{ }\label{algo1a}
\end{minipage}
\begin{minipage}[htb]{0.97\linewidth}
\centering\includegraphics[width=0.83\columnwidth]{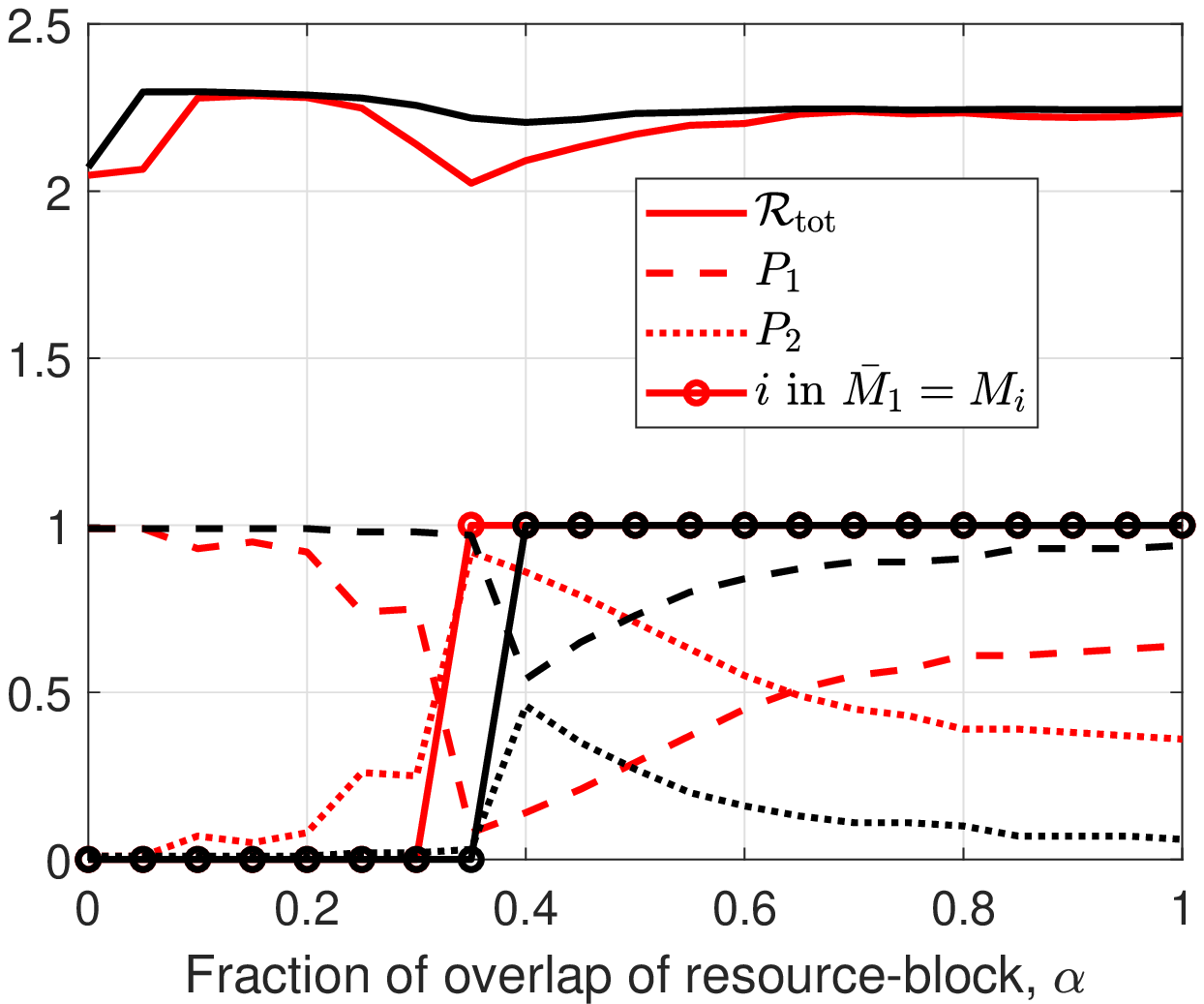}
\subcaption{ }\label{algo1b}
\end{minipage}
\begin{minipage}[htb]{0.97\linewidth}
\centering\includegraphics[width=0.83\columnwidth]{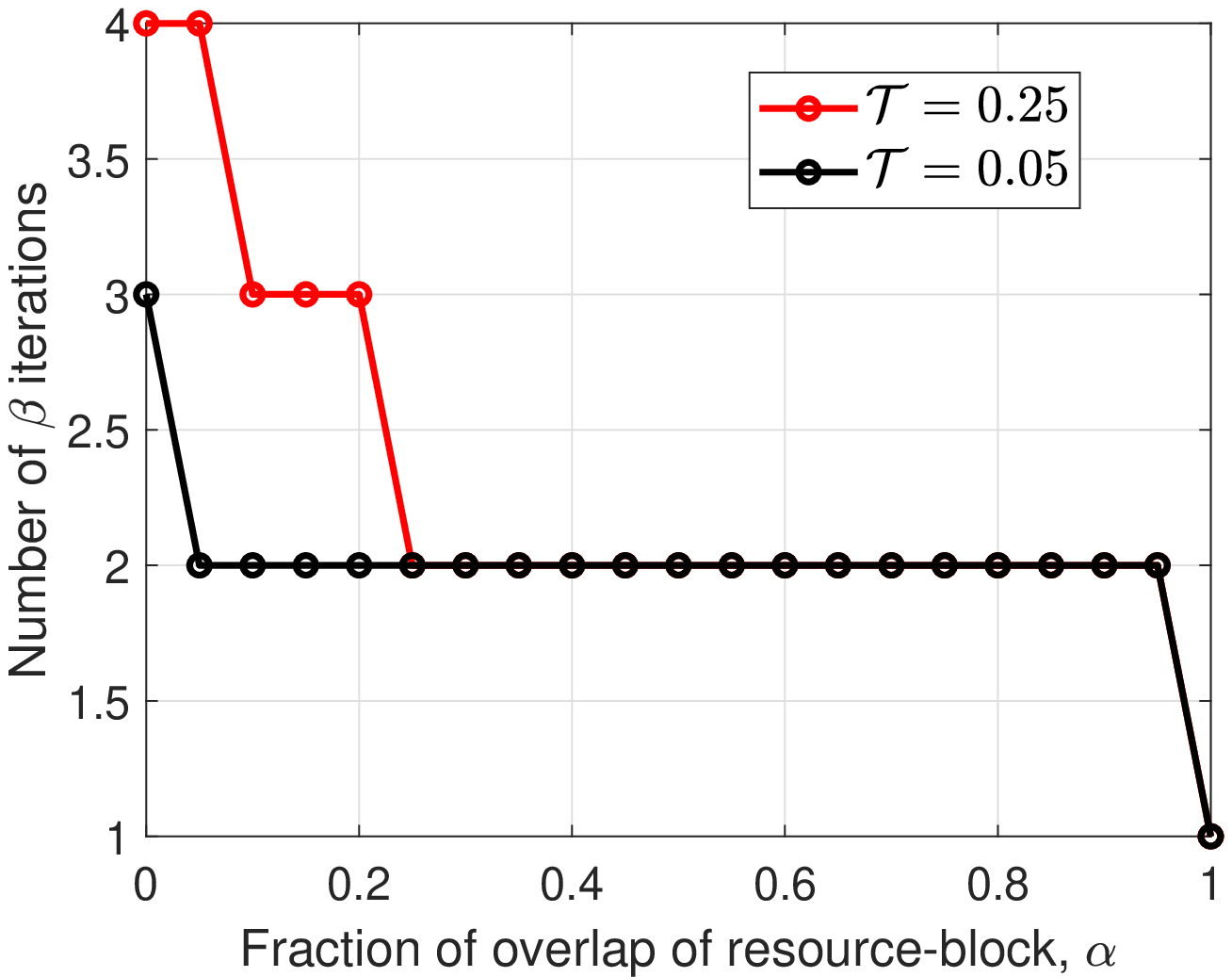}
\subcaption{ }\label{algo1c}
\end{minipage}
\begin{minipage}[htb]{0.97\linewidth}
\centering\includegraphics[width=0.83\columnwidth]{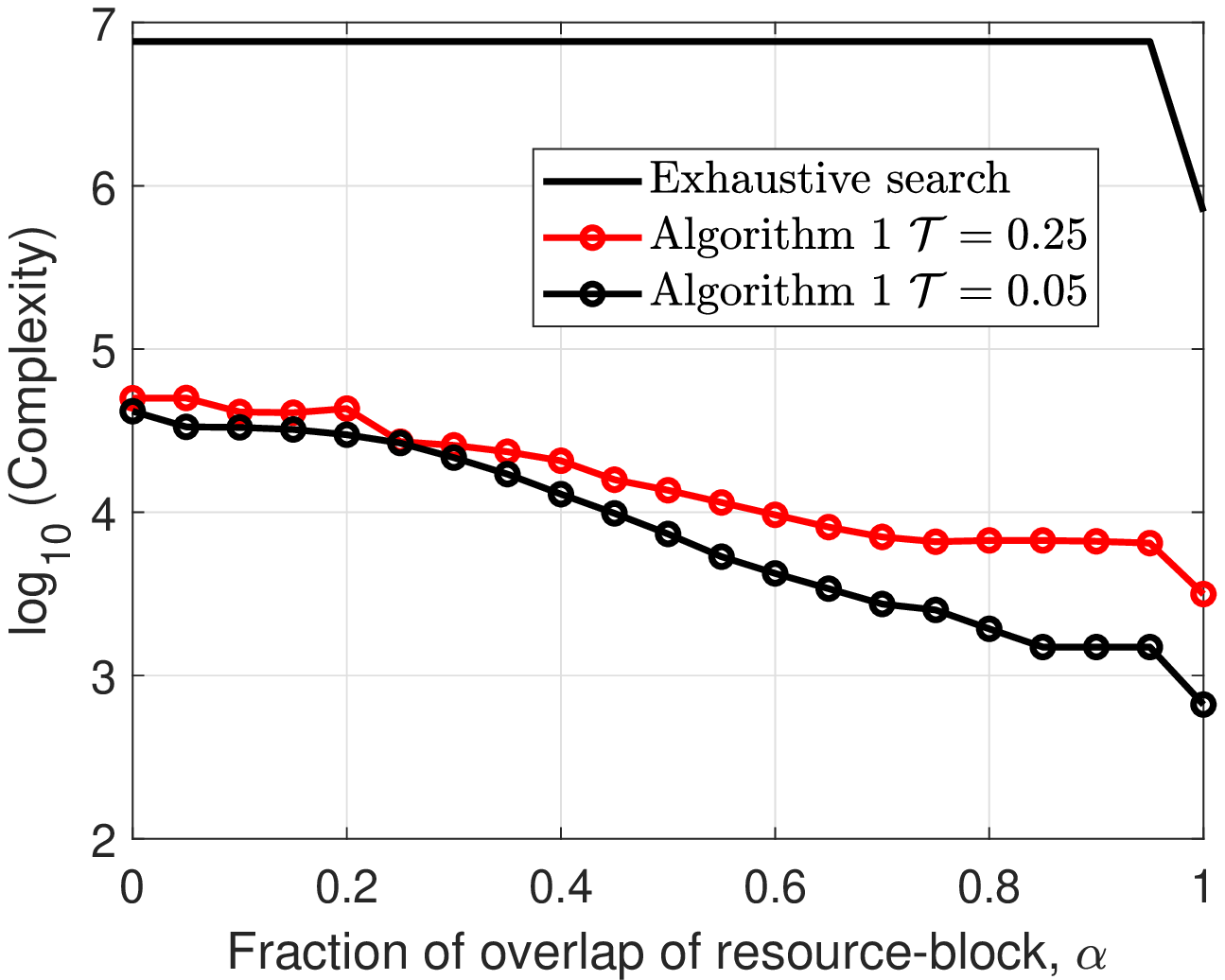}
\subcaption{ }\label{algo1d}
\end{minipage}
\caption{Using \textbf{Algorithm 1} to solve $\mathcal{P}1$ with increasing $\alpha$. Red curves are for $\mathcal{T}=0.25$ and black curves are for $\mathcal{T}=0.05$.}\label{algo1Results}
\end{figure}

Fig. \ref{algo1Results} plots the impact of increasing $\alpha$ on different elements when the cell sum rate is maximized subject to a TMT constraint, i.e., $\mathcal{P}1$. We consider two different values of TMT, $\mathcal{T}=0.05$ and $\mathcal{T}=0.25$. Fig. \ref{algo1a} plots the individual UE and cell sum rates. As has been mentioned, \textbf{Algorithm 1} is used to obtain the resource allocation; hence $\mathcal{R}_2$ attains the TMT. For a given $\mathcal{T}$, we observe in Fig. \ref{algo1b} that upto {$\alpha=0.3$ for $\mathcal{T}=0.25$ ($\alpha=0.35$ for $\mathcal{T}=0.05$)}, $P_2$ slowly increases because of the increasing $\mathcal{I}(\alpha,\beta)$ and consequent intracell interference requiring higher power by $\text{UE}_2$ to achieve TMT. Corresponding to this range of $\alpha$, there is first an increase and then a decrease in $\mathcal{R}_1$ (and therefore $\mathcal{R}_{\rm tot}$) although $P_1$ decreases slowly. The initial increase in rate is attributed to the more significant impact of $\text{BW}_1$ at first; however, after the local optimum, the impact of lower $P_1$ and higher $\mathcal{I}(\alpha, \beta)$, contributing to lower power of the message of interest and higher intracell interference, takes over and causes a degradation in rate as $\alpha$ increases. 

After {$\alpha=0.3$ for $\mathcal{T}=0.25$ ($\alpha=0.35$ for $\mathcal{T}=0.05$)}, we observe from Fig. \ref{algo1b} that, there is a switch in the decoding technique from $\bar{M}_1=M_0$ to $\bar{M}_1=M_1$, i.e., from $\text{UE}_1$ treating the message of $\text{UE}_2$ as noise to decoding it. This switch also corresponds to a sudden increase in $P_2$ leaving behind less power for $\text{UE}_1$'s message. However, we still observe an increase in $\mathcal{R}_1$ because decoding and removing $\text{UE}_2$'s message before decoding its own improves $\text{UE}_1$'s performance. This need for $\text{UE}_1$ to decode $\text{UE}_2$'s message scaled by $\mathcal{I}(\alpha, \beta)$ is also why there is a spike in $P_2$ when the decoding technique switches. As $\alpha$ grows, $P_2$ decreases as $\mathcal{I}(\alpha, \beta)$ grows so it is easier for $\text{UE}_1$ to decode the weak UE's message and because lower $P_2$ is required by $\text{UE}_2$ which also has larger bandwidth as the overlap $\alpha$ grows. Hence, as $\alpha$ grows after the switch in $\bar{M}_1$, $\mathcal{R}_{\rm tot}$ grows with $\alpha$.

We also observe that lower $\mathcal{T}$ corresponds to higher $\mathcal{R}_{\rm tot}$ as there are more resources available for $\text{UE}_1$ to maximize its throughput with. Additionally, we observe a range of $\alpha$ that outperforms traditional NOMA in terms of $\mathcal{R}_\mathrm{tot}$ and thus there exists an optimum $\alpha \neq 1$ that maximizes cell sum rate subject to a TMT constraint. Note that this is in line with partial-NOMA outperforming traditional NOMA in the rate region abstraction as unconstrained cell sum rate maximization is equivalent to $\mathcal{P}1$ with $\mathcal{T}=0$. {It ought to be noted that the values of $\alpha$ (including the optimum) that outperform traditional NOMA in terms of $\mathcal{R}_\mathrm{tot}$ occur in the region where $\bar{M}_1=M_0$ (i.e., the message of the weak UE is treated as noise by the strong UE). This highlights the important role that FSIC plays in the superiority of partial-NOMA.}

In Fig. \ref{algo1c}, we plot the number of iterations of $\beta$ required by the algorithm. Note that the algorithm actually goes through one more iteration than the iteration at which the optimum $\beta$ is found for $\alpha<1$ as we terminate the algorithm once $\mathcal{R}_{\rm tot}$ starts decreasing with decreasing $\beta$. For $\alpha=1$, there is only one possible value of $\beta$, and therefore, only one iteration. We observe that a higher $\mathcal{T}$ naturally requires more iterations since a lower $\text{BW}_2$ may not be sufficient to meet the TMT or may be consuming too much $P_2$, thereby requiring us to increase $\text{BW}_2$ by reducing $\beta$. Note that the number of iterations required decreases monotonically with $\alpha$ for a given $\mathcal{T}$. This is because, as $\alpha$ increases, $\text{BW}_2$ increases and so the need to decrease $\beta$ is less.


{As mentioned in Section IV, it is impractical to carry out an exhaustive search for each value of $\alpha$ due to the high complexity involved. However, since we have data from the exhaustive searches in Fig. \ref{RatesVsP} for $\alpha$ values of \{0, 0.05, 0.1, 0.2, 0.25, 0.35, 0.5, 0.9, 1\}, we use these to benchmark the performance of the results in Fig. \ref{algo1a}. For these values of $\alpha$ with both TMT values used in Fig. \ref{algo1Results} we find matches with the data from the exhaustive searches. As mentioned in Remark 5, we would like to emphasize that while this is not proof for the optimality of our algorithm, it sheds light on the accuracy of our feasible solution.}

In Fig. \ref{algo1Results}, we search for the transmission rates in $-20 \text{ dB} \leq \theta_i \leq 21 \text{ dB}$ and use step size $\Delta_{\theta}=0.5$. As a result, there are $\hat{\Delta}_{\theta}=83$ choices of $\theta_i$, $i \in \{1,2\}$. For $P_2$, we use step size $\Delta_P=0.01$; hence, there are $\hat{\Delta}_P=101$ choices of $P_2$. $\Delta_{\beta}$ is already defined in Section \ref{4c} so that there are $\hat{\Delta}_{\beta}=11$ choices of $\beta$ when $\alpha<1$ and there is only $\hat{\Delta}_{\beta}=1$ choice of $\beta$ when $\alpha=1$. {While we do not carry out an exhaustive search in Fig. \ref{algo1Results}, we still compare the complexity of our algorithm with that of an exhaustive search as it is constant for the latter.} As has been mentioned, for a fair comparison, the same search space is considered for the exhaustive search.

Fig. \ref{algo1d} shows that the proposed \textbf{Algorithm 1} requires significantly lower complexity than an exhaustive search. Additionally, the complexity of \textbf{Algorithm 1} increases as $\mathcal{T}$ increases due to the larger number of iterations of both $P_2$ and $\beta$ required to achieve TMT and find the optimum. We also observe that overall as $\alpha$ increases, the complexity of \textbf{Algorithm 1} decreases\footnote{{Our complexity curves for \textbf{Algorithm 1} are not very smooth as the grid for $\theta_i$ is not very fine. This sometimes results in a longer search for the resources that allow $\text{UE}_2$ to attain TMT and therefore the decrease in complexity with $\alpha$ is not monotonic (by small amounts). A finer $\theta_i$ grid would result in smoother curves but the price paid would be much higher complexity. As the difference in resource allocation and performance would be marginal, we do not do this to avoid longer computation times.}}. This is in line with the fact that a higher value of $\alpha$ requires a fewer number of iterations of $\beta$ for the algorithm to find the optimum solution. It should also be noted that for both the exhaustive search and \textbf{Algorithm 1}, there is a decrease in complexity at $\alpha=1$ because there is only one value of $\beta$ possible making the search space smaller.

\subsection{Summary of Main Results}
{
We summarize the main results as follows:
\begin{itemize}
\item The coverage of $\text{UE}_2$ decreases monotonically with the overlap $\alpha$. The coverage of $\text{UE}_1$, on the other hand, does not. This is because the impact of increasing $\mathcal{I}(\alpha,\beta)$ is not as trivial due to FSIC decoding.
\item Some choices of resource allocation will result in guaranteed outage for certain $\alpha$, emphasizing the importance of careful resource allocation.
\item With appropriate resource allocation, partial-NOMA results in better coverage than traditional NOMA for both UEs.
\item Traditional SIC is inadequate in the low $\alpha$ regime, while treating the message of the weak UE as noise at the strong UE is inadequate when $\alpha$ is higher.
\item Reducing the overlap $\alpha$ allows the system to support UEs with higher transmission rate requirements. This allows partial-NOMA to serve UEs that traditional NOMA cannot, thereby preventing inefficient spectrum reuse.
\item As anticipated, for a given power and transmission rate allocation, the impact of increased interference with $\alpha$ is lower than that of increased bandwidth.
\item Using \textbf{Algorithm 1}, when $\bar{M}_1=M_0$, we observe a local optimum for $\mathcal{R}_1$ although $P_1$ is decreasing in this range because of the trade off between increasing bandwidth and decreasing SINR (due to lower signal power and higher intracell interference).
\item Using \textbf{Algorithm 1}, after the switch from $\bar{M}_1=M_0$ to $\bar{M}_1=M_1$, $\mathcal{R}_1$ increases with $\alpha$.
\item {We observe partial-NOMA to outperform traditional NOMA in terms of $\mathcal{R}_{\rm tot}$ both in the rate region and using \textbf{Algorithm 1}. This occurs in the low $\alpha$ regime where $\bar{M}_1=M_0$ highlighting the important role that FSIC plays in the superiority of partial-NOMA.
\item An optimum $\alpha<1$ exists given a TMT constraint that maximizes $\mathcal{R}_\mathrm{tot}$.}
\item \textbf{Algorithm 1} is shown to have much lower complexity than an exhaustive search. Its complexity grows with TMT and decreases with $\alpha$.
\end{itemize}}

\section{Conclusion}\label{Conc}
Partial-NOMA is proposed as a technique to strike a balance between the high interference associated with NOMA resulting in low coverage and no spectrum reuse in OMA resulting in low rates. A large downlink two-user network employing partial-NOMA is studied. The nature of the partial overlap allows us to employ receive-filtering to further suppress the interference in a partial-NOMA setup. The received filtering not only suppresses intracell interference but also results in a suppression of intercell interference allowing our setup to have lower intercell interference than both NOMA and OMA. A technique called FSIC decoding is proposed for decoding the partial-NOMA setup. An abstraction of the rate region is studied and compared to that of NOMA and OMA. It is observed that for some values of the overlap $\alpha$, partial-NOMA can outperform NOMA. It is also shown that partial-NOMA allows more flexibility in terms of the achievable individual UE throughput. A problem of maximizing cell sum rate subject to a TMT constraint is formulated. Since the problem is non-convex, the only known solution requires an exhaustive search. An efficient algorithm that finds a feasible solution to the problem is proposed. The complexity of the algorithm is shown to be much lower than an exhaustive search. It is shown that the partial-NOMA setup outperforms NOMA in terms of cell sum rate for {a range of values of} $\alpha$. {Additionally, in this range, there exists an optimum value of $\alpha$ that maximizes the cell sum rate for a given TMT constraint. Furthermore, it is observed that the range of $\alpha$ which results in superior cell sum rate to traditional NOMA corresponds to $\bar{M}_1=M_0$, highlighting the role of FSIC in the superiority of partial-NOMA}. It is also shown that while NOMA cannot support UEs with high transmission rate requirements, partial-NOMA can. Instead of allocating an entire resource-block to such UEs via OMA, these UEs ought to be served via partial-NOMA to reuse the spectrum efficiently. The work in this paper studies a two-user setup. An important direction for future work is to study partial-NOMA in an $N$-user setup, where $N$ is general. In this paper, we have considered downlink transmissions; investigating partial-NOMA in the uplink is also an important and interesting direction.

\appendices

\bibliographystyle{IEEEtran}
\bibliography{References}

\end{document}